\documentclass{article}

    \PassOptionsToPackage{numbers, compress}{natbib}

\usepackage[final]{neurips_2020}

\usepackage[utf8]{inputenc} 
\usepackage[T1]{fontenc}    
\usepackage{hyperref}       
\usepackage{url}            
\usepackage{booktabs}       
\usepackage{amsfonts}       
\usepackage{nicefrac}       
\usepackage{microtype}      

\usepackage{graphicx}
\usepackage{amsmath,amssymb}
\usepackage{xcolor}
\usepackage{bm}
\usepackage{subcaption}
\usepackage[title]{appendix}
\usepackage{wrapfig}
\usepackage{todonotes}

\DeclareMathOperator{\E}{\mathbb{E}}
\DeclareMathOperator{\bx}{\mathbf{x}}
\DeclareMathOperator{\bc}{\mathbf{c}}
\DeclareMathOperator{\by}{\mathbf{y}}
\DeclareMathOperator{\bz}{\mathbf{z}}
\DeclareMathOperator{\bs}{\mathbf{s}}

\newcommand{\vect}[1]{\boldsymbol{\mathbf{#1}}}

\newcommand\blfootnote[1]{%
  \begingroup
  \renewcommand\thefootnote{}\footnote{#1}%
  \addtocounter{footnote}{-1}%
  \endgroup
}

\title{A Spectral Energy Distance\\ for Parallel Speech Synthesis}

\author{Alexey A. Gritsenko$^{* \dagger}$ \quad Tim Salimans$^{*}$  \quad Rianne van den Berg  \\ \textbf{Jasper Snoek}  \quad \textbf{Nal Kalchbrenner} 
\\
\{agritsenko,salimans,riannevdberg,jsnoek,nalk\}@google.com
\\
Google Research}

\newcommand{\alexey}[1]{}
\renewcommand{\alexey}[1]{{\color{orange} \textbf{Alexey}: {#1}}}

\newcommand{\tim}[1]{}
\renewcommand{\tim}[1]{\todo[inline,color=olive]{\textbf{Tim}: {#1}}}

\newcommand{\rianne}[1]{}
\renewcommand{\rianne}[1]{\todo[inline,color=blue]{\textbf{Rianne}: {#1}}}

\newcommand{\jasper}[1]{}
\renewcommand{\jasper}[1]{\todo[inline,color=magenta]{\textbf{Jasper}: {#1}}}

\newcommand{\nal}[1]{}
\renewcommand{\nal}[1]{\todo[inline,color=teal]{\textbf{Nal}: {#1}}}

\begin{document}

\maketitle

\blfootnote{\kern-1.7em$^*$Equal contribution. $^\dagger$ Work completed as a Google AI resident.}

\begin{abstract}
Speech synthesis is an important practical generative modeling problem that has seen great progress over the last few years, with likelihood-based autoregressive neural models now outperforming traditional concatenative systems. A downside of such autoregressive models is that they require executing tens of thousands of sequential operations per second of generated audio, making them ill-suited for deployment on specialized deep learning hardware. Here, we propose a new learning method that allows us to train highly parallel models of speech, without requiring access to an analytical likelihood function. Our approach is based on a generalized energy distance between the distributions of the generated and real audio. This \emph{spectral energy distance} is a proper scoring rule with respect to the distribution over magnitude-spectrograms of the generated waveform audio and offers statistical consistency guarantees. The distance can be calculated from minibatches without bias, and does not involve adversarial learning, yielding a stable and consistent method for training implicit generative models. Empirically, we achieve state-of-the-art generation quality among implicit generative models, as judged by the recently-proposed cFDSD metric. When combining our method with adversarial techniques, we also improve upon the recently-proposed GAN-TTS model in terms of Mean Opinion Score as judged by trained human evaluators.
\end{abstract}

\section{Introduction}
Text-to-speech synthesis (TTS) has seen great advances in recent years, with neural network-based methods now significantly outperforming traditional concatenative and statistical parametric approaches \citep{Zen2009, VanDenOord2016}. While autoregressive models such as WaveNet \citep{VanDenOord2016} or WaveRNN \citep{Kalchbrenner2018} constitute the current state of the art in speech synthesis, their sequential nature is often seen as a drawback. They generate only a single sample at a time, and since audio is typically sampled at a frequency of 18kHz to 44kHz this means that tens of thousands of sequential steps are necessary for generating a single second of audio. The sequential nature of these models makes them ill-suited for use with modern deep learning hardware such as GPUs and TPUs that are built around parallelism.

At the same time, parallel speech generation remains challenging. Existing likelihood-based models either rely on elaborate distillation approaches \citep{Ping2018,VanDenOord2017}, or require large models and long training times \citep{Prenger2019, Kumar2019}. Recent GAN-based methods provide a promising alternative to likelihood-based methods for TTS \citep{Binkowski2019, Kumar2019}. Although they do not yet match the speech quality of autoregressive models, they are efficient to train and employ fully-convolutional architectures, allowing for efficient parallel generation. However, due to their reliance on adversarial training, they can be difficult to train.


To address these limitations we propose a new training method based on the \textit{generalized energy distance} \citep{Gneiting2007, Sejdinovic2013, Shen2018}, which enables the learning of implicit density models without the use of adversarial training or requiring a tractable likelihood. Our method minimizes a multi-resolution spectrogram loss similar to previous works \citep{Wang2019, Yamamoto2020, Engel2020,openai_jukebox}, but includes an additional \emph{repulsive term} that encourages diverse samples and provides a statistical consistency guarantee. As a result, our models enjoy stable training and rapid convergence, achieving state-of-the-art speech quality among implicit density models.

In addition to demonstrating our proposed energy distance on the speech model of \citet{Binkowski2019}, we further propose a new model for generating audio using an efficient overlap-add upsampling module. The new model is faster to run, while still producing high quality speech. Finally, we show that our proposed energy distance can be combined with GAN-based learning, further improving on either individual technique. An open source implementation of our generalized energy distance is available at \url{https://github.com/google-research/google-research/tree/master/ged_tts}.

\section{Related work on speech synthesis}
\label{section:related-work}
Our task of interest is synthesizing speech waveforms conditional on intermediate representations such as linguistic and pitch features, as usually provided by a separate model in a 2-step process. Here we briefly review the related literature on this problem.

\paragraph{Autoregressive models.} \citet{VanDenOord2016} proposed WaveNet, an autoregressive model that produces high-fidelity speech by directly generating raw waveforms from the input features. WaveNet is trained by maximizing the likelihood of audio data conditional on the aforementioned linguistic and pitch features.
While WaveNet's fully convolutional architecture enables efficient training on parallel hardware, at inference time the model relies on an autoregressive sampling procedure, generating the waveform one sample at a time. This necessitates tens of thousands of sequential steps for generating a single second of audio, making it ill-suited for real-time production deployment. These limitations were partially alleviated by \citet{Kalchbrenner2018}. While still autoregressive, using a single-layer recurrent neural network, weight sparsification and custom kernels, their WaveRNN model achieves faster-than-realtime on-device synthesis.

\paragraph{Probability density distillation.}
Parallel~WaveNet \citep{VanDenOord2017} and ClariNet \citep{Ping2018} used a trained autoregressive model such as WaveNet as a teacher network \textit{distilling} its knowledge into a non-autoregressive likelihood student model that is trained to minimize the Kullback-Liebler (KL) divergence between student and teacher. Both approaches rely on an Inverse-Autoregressive Flow (IAF; \citet{Kingma2016}) as a student network.
The IAF is structured in such a way that, given a set of latents, the corresponding observables can be generated efficiently in parallel.

While the methods of \citet{Ping2018} and \citet{VanDenOord2017} differ in the choice of distributions used in their models, they both found that optimizing the KL-divergence alone was insufficient for obtaining high-quality generations, and required careful regularization and auxiliary losses for the student models to converge to a good solution.

\paragraph{Flow-based models.}
To avoid having to use a two-stage training pipeline required by the distillation approaches, FloWaveNet \citep{Kim2018} and WaveGlow \citep{Prenger2019} propose directly training a convolutional flow model based on the architectures of RealNVP \citep{Dinh2016} and Glow \citep{Kingma2018} respectively. These models can be trained using maximum likelihood, and approach the speech quality of WaveNet or its distillations. However, due to the limited flexibility of their coupling layers (the building blocks used to construct invertible models) flow-based approaches tend to require large networks. Indeed, both WaveGlow and FloWaveNet have $\approx 100$ convolutional layers, and are slow to train \citep{Wang2019}.

Concurrently to our work \citet{Ping2019} have made great progress towards reducing the size of flow-based TTS models. Their WaveFlow model achieves high quality speech synthesis using $128$ sequential steps (more than two orders of magnitude fewer than a fully autoregressive model) while maintaining a small parameter footprint.

\paragraph{Implicit generative models.} 
To date, Generative Adversarial Networks (GANs; \citet{Goodfellow2014}) are mostly applied to image generation, where they are known to produce crisp, high-quality images using fully convolutional architectures and relatively small models \citep{Brock2018}. Their application to speech synthesis may provide an alternative to probability distillation that has the potential to match it in terms of speech quality. Three recent approaches, MelGAN \citep{Kumar2019}, GAN-TTS \citep{Binkowski2019} and Parallel WaveGAN \citep{Yamamoto2020}, made significant progress in this direction. While still not matching WaveNet or ClariNet in speech quality, these works have proven the feasibility of solving TTS with implicit generative models. However, due to the reliance on adversarial learning, GANs can still be difficult to train. MelGAN and GAN-TTS rely on carefully chosen regularization and an ensemble of discriminators to achieve stable training, and MelGAN and Parallel WaveGAN use various auxiliary losses. This training difficulty also limits the available model choice, as only certain types of models (e.g.\ with batch normalization and spectral normalization) are known to be trainable.

\section{Maximum Mean Discrepancy and Energy Distance}
\label{section:mmd}
Before we present our proposed training method in Section~\ref{section:spectral-ged}, we briefly review previous work on learning implicit generative models, and we introduce the primitives on which our method is built. 

Although GANs have recently become the dominant method for training implicit generative models without tractable likelihood, another popular approach to learning these types of models is a class of methods based on minimizing \emph{Maximum Mean Discrepancy} (MMD), defined as 
\begin{equation}
    D_{\text{MMD}}(p|q) = \sup_{f \in \mathcal{H}}\left[ \E_{\bx \sim p(\bx)}[f(\bx)] - \E_{\by \sim q(\by)}[f(\by)]\right],
    \label{eq:mmd}
\end{equation}
where $f$ is a critic function which is constrained to a family of functions $\mathcal{H}$ \citep[see][]{Gretton2012}, and $p(\bx)$ and $q(\by)$ are the data and model distributions respectively. When $\mathcal{H}$ is given by a family of neural network discriminators, MMD becomes very similar to GANs, as explained by \cite{Arjovsky2017}. The main difference is that for MMD the maximization over $f \in \mathcal{H}$ is assumed to be analytically tractable, while GANs maximize over $f$ approximately by taking a few steps of stochastic gradient descent. The benefit of exact optimization is that MMD methods are provably stable and consistent, unlike GANs, although this comes at the cost of more restrictive critic families $\mathcal{H}$.

\citet{Gretton2012} show that exact optimization is indeed possible if $\mathcal{H}$ is chosen to be a reproducing kernel Hilbert space (RKHS). In that case, there exists a kernel function $k \in \mathcal{H}$ such that every critic $f \in \mathcal{H}$ can be expressed through its inner product with that kernel:
\begin{equation}
    f(\bx) = \langle f, k(\cdot,\bx) \rangle_{\mathcal{H}} = \sum_{i} \alpha_{i}k(\bx,\bx_{i}).
\end{equation}
In other words, $f$ is constrained to be a weighted sum of basis functions $k(\bx,\bx_{i})$, with weights $\alpha$. Exact optimization over $\alpha$ then gives the following expression for the squared MMD:
\begin{equation}
    D_{\text{MMD}}^{2}(p|q) = \E[ k(\bx,\bx') + k(\by,\by') - 2 k(\bx,\by) ],
    \label{eq:mmd2}
\end{equation}
where $\bx,\bx' \sim p(\bx)$ and  $\by,\by' \sim q(\by)$ are independent samples from $p$ and $q$.

Since (\ref{eq:mmd2}) only depends on expectations over $q$ and $p$ it can be approximated without bias using Monte Carlo sampling. If our dataset contains $N$ samples from $p(\bx)$ and we draw $M$ samples from our model $q(\by)$, this gives us the following stochastic loss function \citep{Gretton2012}:
\begin{equation}
    L(q) = \frac{1}{N(N-1)} \sum_{n \neq n'} k(\bx_{n},\bx_{n'}) + \frac{1}{M(M-1)} \sum_{m \neq m'} k(\by_{m},\by_{m'}) - \frac{2}{MN} \sum_{n=1}^{N}\sum_{m=1}^{M} k(\bx_n,\by_m).
    \label{eq:mmd3}
\end{equation}

Loss functions of this type were used by \cite{Dziugaite2015, Li2015} and \cite{Bouchacourt2016} to train generative models without requiring a tractable likelihood function.

An alternative view on MMD methods is in terms of \emph{distances}. As explained by \citet{Sejdinovic2013}, the kernel $k(\cdot,\cdot)$ of a RKHS $\mathcal{H}$ induces a distance metric $d(\cdot,\cdot)$ via
\begin{equation}
    d(\bx,\by) = \frac{1}{2}(k(\bx,\bx) + k(\by,\by) - 2k(\bx,\by)).
    \label{eq:dist}
\end{equation}
Assuming that $k(\bx,\bx) = k(\by,\by) = c$ with $c$ being a constant, equation (\ref{eq:mmd2}) can equivalently be expressed in terms of this distance:
\begin{equation}
    D_{\text{MMD}}^{2}(p|q) = D_{\text{GED}}^{2}(p|q) = \E[ 2 d(\bx,\by) - d(\bx,\bx') - d(\by,\by')],
    \label{eq:mmd_dist}
\end{equation}
which is known as the \emph{generalized energy distance} \citep[GED; see e.g.][]{Sejdinovic2013, Shen2018, Salimans2018}.

In most practical applications of generative modeling, such as speech synthesis, we are interested in learning \emph{conditional} distributions $q(\bx|\bc)$ using examples $\bx_{i},\bc_{i}$ from the data distribution $p$. In such cases we usually only have access to a single example $\bx_{i}$ for each unique conditioning variable $\bc_{i}$. This means that we cannot evaluate the term $\E[d(\bx,\bx')]$ in (\ref{eq:mmd_dist}). 
However, this term only depends on the data distribution $p$ and not on our generative model $q$, so it can be dropped during training.
The training loss then becomes
\begin{equation}
    L_{\text{GED}}(q) = \E[ 2 d(\bx,\by) - d(\by,\by')],
    \label{eq:energy_score}
\end{equation}
with $\by,\by' \sim q(\cdot|\bc)$ independent samples from our model, conditioned on the same features $\bc$. This type of loss was studied by \citet{Gneiting2007} under the name \emph{energy score}. They find that (\ref{eq:energy_score}) is a \emph{proper scoring rule}, i.e.\ it can lead to a statistically consistent learning method, if the distance metric $d(\cdot, \cdot)$ is negative definite. This result is more general than the consistency results for MMD, and also allows for the use of distances that do not correspond to reproducing kernel Hilbert spaces. We make use of this result for deriving our proposed learning method, which we present in Section~\ref{section:spectral-ged}.

\section{A generalized energy distance based on spectrograms}
\label{section:spectral-ged}
We require a method to learn generative models that can sample speech in a small number of parallel steps, without needing access to a tractable likelihood function. The method we propose here achieves this by computing a \emph{generalized energy distance}, or \emph{energy score}, between simulated and real data, and minimizing this loss with respect to the parameters of our generative model. Here, we assume that our dataset consist of $N$ examples of speech $\bx_{i}$, labeled by textual or linguistic features $\bc_{i}$. Our generative model is then a deep neural network that takes a set of Gaussian noise variables $\bz_{i}$, and maps those to the audio domain as $\by_{i} = f_{\theta}(\bc_{i}, \bz_{i})$, with $\theta$ the parameters of the neural network. This implicitly defines a distribution $q_{\theta}(\by|\bc)$ of audio $\by$ conditional on features $\bc$.

Given a minibatch of $M$ examples $\{\bx_i, \bc_i\}_{i=1}^M$, we use our model to generate two independent samples $\by_{i} = f_{\theta}(\bc_{i}, \bz_{i})$, $\by_{i}' = f_{\theta}(\bc_{i}, \bz_{i}')$ corresponding to each input feature $\bc_{i}$, using two independently sampled sets of noise variables $\bz_{i}, \bz_{i}'$. We then calculate the resulting minibatch loss as
\begin{equation}
L^{*}_{\text{GED}}(q) = \sum_{i=1}^{M} 2d(\bx_{i},\by_{i}) - d(\by_{i},\by_{i}'),
\label{eq:energy_score_mini}
\end{equation}
where $d(\cdot, \cdot)$ is a distance metric between samples. The minibatch loss $L^{*}_{\text{GED}}(q)$ is an unbiased estimator of the energy score (\ref{eq:energy_score}), and minimizing it will thus minimize the generalized energy distance between our model and the distribution of training data, as discussed in Section \ref{section:mmd}.

In practice the performance of the energy score strongly depends on the choice of metric $d(\cdot, \cdot)$. When generating high-dimensional data, it is usually impossible to model all aspects of the data with high fidelity, while still keeping the model $q_{\theta}(\by|c)$ small enough for practical use. We thus have to select a distance function that emphasizes those features of the generated audio that are most important to the human ear. This is similar to how GANs impose a powerful visual inductive bias when modeling images using convolutional neural network discriminators. Following the literature on \emph{speech recognition} \citep{Kingsbury1998, Amodei2016}, we thus define our distance function over \emph{spectrograms} $\bs^{k}(\bx_{i})$, where a spectrogram is defined as the magnitude component of the short-time Fourier transform (STFT) of an input waveform, $\left|\text{STFT}_{k}(\bx_{i})\right|$, where $k$ is the frame-length used in the STFT. Following \citet{Engel2020} we combine multiple such frame-lengths $k$ into a single multi-scale spectrogram loss. Our distance function to be used in the generalized energy distance then becomes
\begin{equation}
d(\bx_{i},\bx_{j}) = \sum_{k \in [2^6, \ldots, 2^{11}]}\sum_{t} ||\bs^{k}_{t}(\bx_{i}) - \bs^{k}_{t}(\bx_{j})||_{1} + \alpha_{k}||\log \bs^{k}_{t}(\bx_{i}) - \log \bs^{k}_{t}(\bx_{j})||_{2},
\label{eq:combined_dist}
\end{equation}
where we sum over a geometrically-spaced sequence of window-lengths between 64 and 2048, and where $\bs^{k}_{t}(\bx_{i})$ denotes the $t$-th timeslice of the spectrogram of $\bx_{i}$ with window-length $k$. The weights $\alpha_{k}$ of the L2 components of the distance are discussed in Appendix~\ref{appendix:proper}. As we show there, the analysis of \citet[Theorem 5.1]{Gneiting2007} can be used to show that this choice makes (\ref{eq:energy_score_mini}) a strictly proper scoring rule for learning $q_{\theta}(\bx|\bc)$ with respect to the ground-truth conditional distribution over spectrograms, meaning that $L_{\text{GED}}(q) > L_{\text{GED}}(p)$ for any $q(\bs^{k}(\bx)|\bc) \neq p(\bs^{k}(\bx)|\bc)$. Minimizing this easily computable loss, we thus obtain a stable and statistically consistent learning method. 

\subsection{Why we need the repulsive term}
Spectrogram-based losses are popular in the literature on audio generation. For example, the probability distillation methods ClariNet \citep{Ping2018} and Parallel~WaveNet \citep{VanDenOord2017} minimize the distance between spectrogram magnitudes of real and synthesized speech in addition to their main distillation loss; and \citet{Pandey2019} use a spectrogram-based loss for speech enhancement. Multi-resolution spectrogram losses like ours were used previously by \citet{Wang2019} and \citet{Yamamoto2020} for speech synthesis, and by \citet{Engel2020} and \citet{openai_jukebox} for music generation. The main difference between these approaches and our generalized energy distance (Equation~\ref{eq:combined_dist}) is the presence of a \emph{repulsive term} between generated data in our training loss, $-d(\by_{i},\by_{i}')$, in addition to the attractive term between generated data and real data, $d(\bx_{i},\by_{i})$.

\begin{figure}[htb]
\centering
\begin{subfigure}{.49\textwidth}
  \centering
  \includegraphics[width=0.97\linewidth]{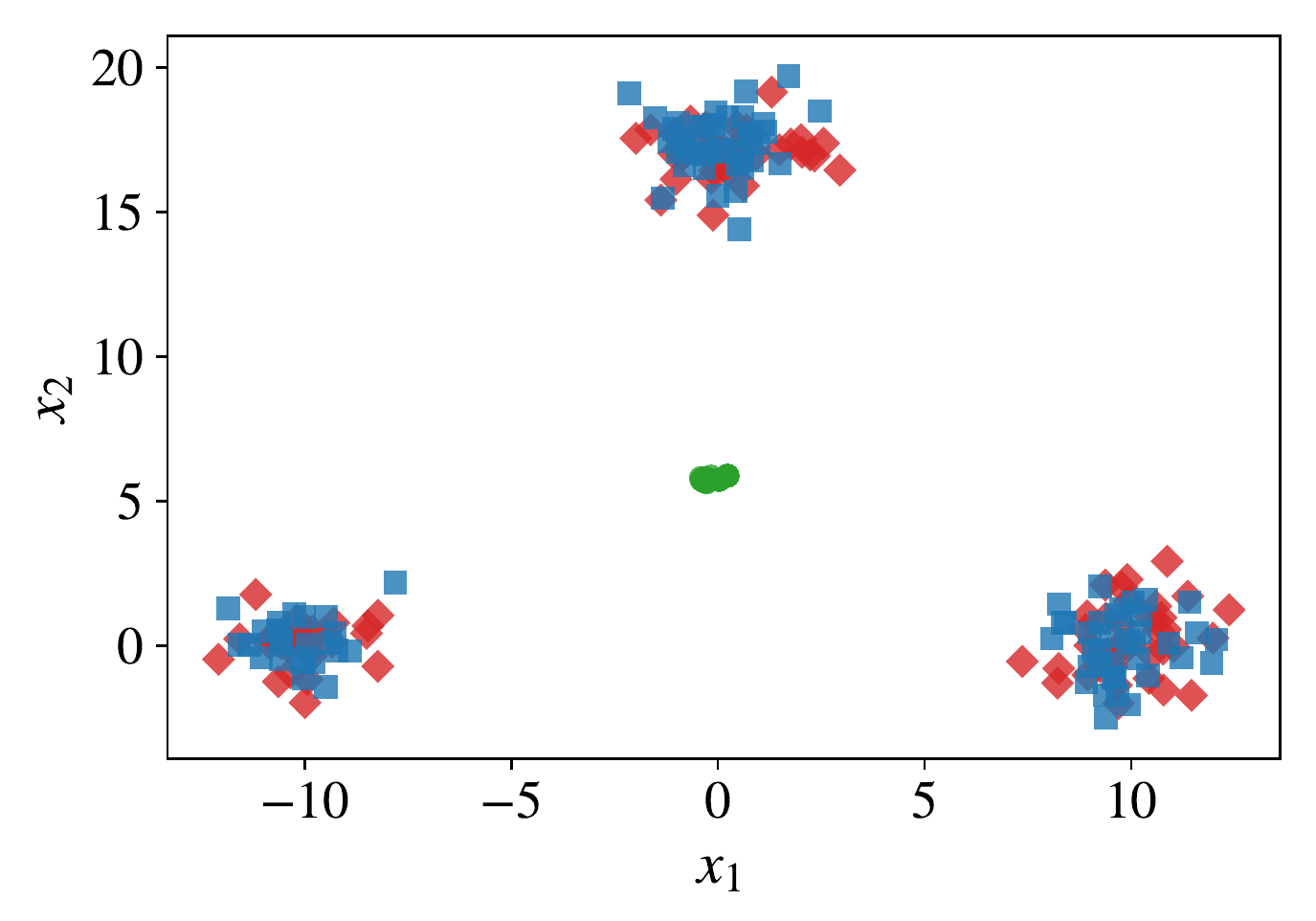}
  \caption{Samples from a two-dimensional Gaussian mixture model with three components.}
  \label{fig:repsulse_mix}
\end{subfigure}\hfill%
\begin{subfigure}{.49\textwidth}
  \centering
  \includegraphics[width=1.\linewidth]{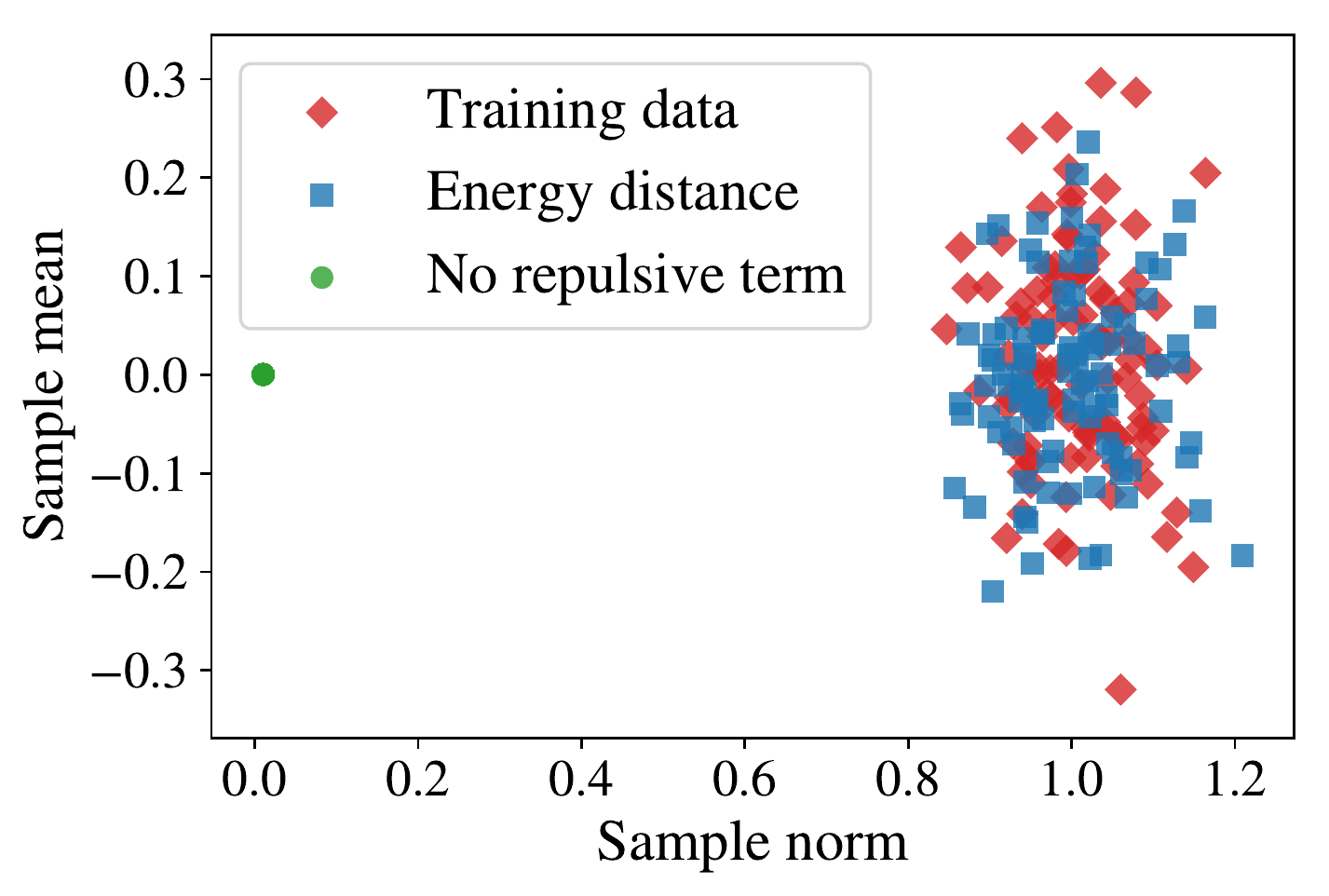}
  \caption{Samples $\mathbf{x}$ from a single 100-dim Gaussian, with $||\mathbf{x}||_{2}$ on the x-axis and $\sum_{i}^{n} \mathbf{x}_{i} / n$ on the y-axis.}
  \label{fig:repulse_high_dim}
\end{subfigure}
\caption{Samples from models trained by minimizing the energy distance (blue) or the more commonly used loss without repulsive term (green), and comparing to samples from the training data (red). Samples from the energy distance trained model are representative of the data, and all sampled points lie close to training examples. Samples from the model trained without repulsive term are not typical of training data. A notebook to reproduce these plots is included in our github repository.}
\label{fig:whyrepulse}
\end{figure}

The presence of the repulsive term is necessary for our loss to be a proper scoring rule for learning the conditional distribution of audio given linguistic features. Without this term, generated samples will collapse to a single point without trying to capture the full distribution. For many purposes like speech and music synthesis it might be argued that a single conditional sample is all that is needed, as long as it is a good sample. Unfortunately the standard loss \textit{without} the repulsive term also fails at this goal, as shown in Figure~\ref{fig:whyrepulse}. If the conditional distribution of training data is multi-modal, regression losses without repulsive term can produce samples that lie far away from any individual mode (Figure~\ref{fig:repsulse_mix}). Even if the conditional distribution of training data is unimodal, such losses will tend to produce samples that are atypical of training data when applied in high dimensions (Figure~\ref{fig:repulse_high_dim}).

In Section~\ref{section:experiments} we perform ablation experiments to further examine the role of the repulsive term for our specific application of speech synthesis. There, we show that this term is critical for achieving optimal performance in practice.

\section{Model and training procedure}
\label{section:model:architecture}
The models we train using the loss function we derived in Section~\ref{section:spectral-ged} consist of deep neural networks that map noise variables to the audio domain, conditioned on linguistic features, that is $\by_{i} = f_{\theta}(\bc_{i}, \bz_{i})$. This is similar to how conditional generator networks are usually parameterized in GANs, see e.g.~BigGAN~\citep{Brock2018} for the analogous case where images $\by$ are generated from noise $\bz$ and class labels $\bc$. For the generator network $f_{\theta}$ we explore 2 different choices:

\paragraph{Simplified GAN-TTS generator} To clearly demonstrate the effect that using the generalized energy distance has on model training, we attempt to control for other sources of variation by using a generator architecture nearly identical to that of GAN-TTS~\citep{Binkowski2019}. Specifically, we use a deep 1D convolutional residual network \citep{He2016} consisting of $7$ residual blocks (see Figure~\ref{fig:spectre-architecture} of the Appendix). 
Compared to the generator of GAN-TTS, we simplify the model by removing the Spectral Normalization \citep{Miyato2018a} and output Batch Normalization \citep{Szegedy2016}, which we empirically find to be either unnecessary or hurting model performance.

\paragraph{Inverse STFT architecture}
To experiment with the wider choice in generative models allowed by our training method, we additionally explore a model that makes use of the Short Time Fourier Transform (STFT) representation that is prevalent in audio processing, and which we also used to define the energy distance we use for training. This model takes in the features and noise variables, and produces an intermediate representation $\text{stft}_{i} = f_{\theta}(\bc_{i}, \bz_{i})$ which we interpret as representing a STFT of the waveform $\by_{i}$ that is to be generated. Here, $f_{\theta}$ consists of a stack of standard ResNet blocks that is applied without any upsampling, and is therefore faster to run than our simplified GAN-TTS generator. We then linearly project $\text{stft}_{i}$ to the waveform space by applying an inverse STFT transformation, thereby upsampling $120\times$ in one step. The final output of this model is thus a raw waveform, similar to the (simplified) GAN-TTS model. 
Further details on this architecture are given in Appendix~\ref{appendix:istft}.

\paragraph{Training procedure}
\setlength{\belowcaptionskip}{-.8\baselineskip}

\begin{wrapfigure}{r}{0.55\textwidth}
  \centering
    \includegraphics[width=0.55\textwidth]{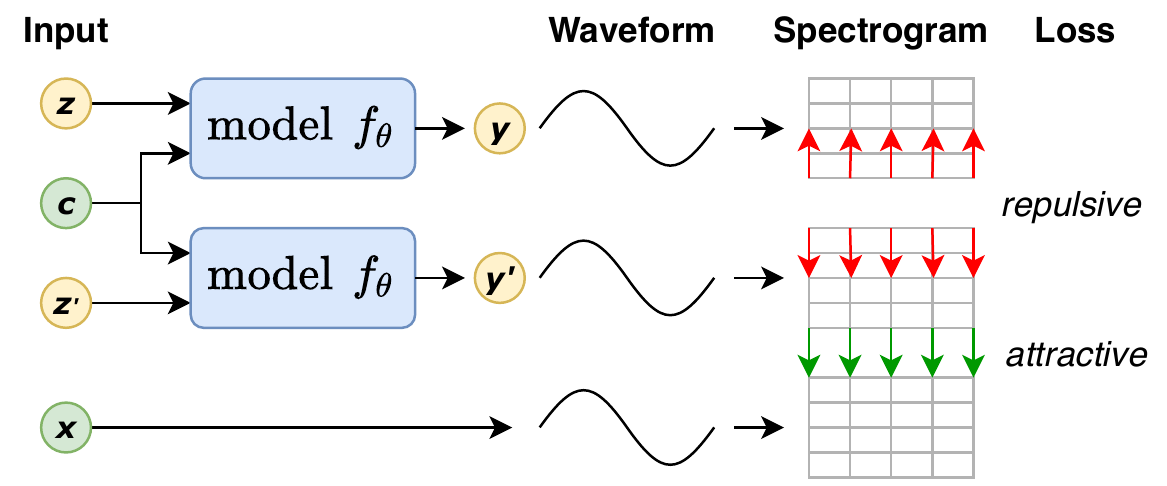}
    \caption{Visual depiction of our training process.}
    \label{fig:training}
\end{wrapfigure}
All of our models are trained on Cloud TPUs v3 with hyper-parameters as described in Table~\ref{appendix:table:spectre-hyperparamters} of the Appendix. For each training example we generate two independent batches of audio samples from our model, conditioned on the same features, which are then used to compute our training loss. Our model parameters are updated using Adam~\citep{Kingma2014}. Figure~\ref{fig:training} explains this training procedure visually.

\section{Data}
\label{section:model:datset}
Our TTS models are trained on a single-speaker North American English dataset, consisting of speech data spoken by a professional voice actor. The data consists of approximately sixty thousand utterances with durations ranging approximately from $0.5$~seconds to $1$~minute, and their corresponding \textit{aligned} linguistic features and pitch information. Linguistic features encode phonetic and duration information, while the pitch is given by the logarithmic fundamental frequency $\log F_0$. These features, a total of $567$, are used by our models as inputs and provide local conditioning information for generating the raw waveform. At training time, the features are derived from and aligned with the training audio. At test time, all features are predicted by a separate model; we thus never use ground-truth information extracted from human speech when evaluating our models.

To account for differences in utterance duration, during training we sample training examples with probability proportional to their length, and then uniformly sample $2$~second windows and their corresponding features from these utterances. Examples shorter than $2$~seconds are filtered out, leaving a total of $44.6$ hours of speech sampled at 24 kHz. The corresponding conditioning features are provided at 5ms intervals (200 Hz sampling frequency). The trained models are thus tasked with converting linguistic and pitch features into raw audio, while upsampling by a factor of $120$. Our training setup is similar to that of \citet{Binkowski2019}, except that we do not apply any transformations (e.g. a $\mu$-transform) to the waveforms.

\section{Experiments}
\label{section:experiments}
We evaluate our proposed approach to speech synthesis by training 4 different models on the data set described in the previous section:
\begin{enumerate}
    \item The simplified GAN-TTS generator described in Section~\ref{section:model:architecture}, but trained by minimizing our generalized energy distance.
    \item This same model and loss, but leaving out the repulsive term $-d(\by_{i},\by_{i}')$ as in previous works that use spectrogram-based losses.
    \item Our inverse STFT model from Section~\ref{section:model:architecture} trained with the GED loss.
    \item A hybrid architecture where we combine the GED loss with adversarial training using an ensemble of \textbf{un}conditional discriminators. Previous works like GAN-TTS and MelGAN use an ensemble of GAN discriminators with \emph{conditional} discriminators taking in features $\bc$, and \emph{unconditional} discriminators that look only at the generated data $\by$. We hypothesize that our GED loss is a good substitute for the conditional part of the GAN loss, but that using an unconditional discriminator might yet offer additional benefit.
\end{enumerate}

To evaluate our generated speech, we report the (conditional) Fr\'echet Deep Speech Distances (FDSD and cFDSD; \citet{Binkowski2019}) - metrics that judge the quality of synthesized audio samples based on their distance to a reference set. These distances are conceptually similar to the FID (Fr\'echet Inception Distance; \citet{Heusel2017}) commonly used for evaluating GANs of natural images, but differ in that they (\textbf{i}) are computed on the activations of the Deep~Speech~2 \citep{Amodei2016} speech recognition model in place of the activations of the Inception network \citep{Szegedy2016} used in FID; and (\textbf{ii}) are computed for samples with conditioning features that match the reference set in the case of the conditional FDSD (cFDSD). We closely followed \citet{Binkowski2019} in our implementation of the FDSD metrics, but note several differences between the two implementations in Appendix~\ref{appendix:FDSD}. We report these metrics on both the training data as well as a larger validation set. In addition, we also evaluate the quality of the synthesized audio using the Mean Opinion Score (MOS) computed from ratings assigned by trained human evaluators. These samples are generated on an \textit{independent} set of $1000$ out-of-distribution sentences for which no ground-truth data is available. 

We compare our models against a careful re-implementation of GAN-TTS \citep{Binkowski2019}, built with help from the authors. Results are reported in Table~\ref{table:results} and include links to samples from each of our models.
\begin{table}[hbt]
\caption{Mean Opinion Score (MOS) and (conditional)  Fr\'echet Deep Speech Distances \citep{Binkowski2019} (FDSD and cFDSD respectively) for prior work and the proposed approach. Models trained by minimizing our spectral generalized energy distance are indicated with GED. Our proposed generator using inverse STFT upsampling is marked iSTFT. For FDSD and cFDSD we report training scores for comparison to the numbers in \citet{Binkowski2019}, as well as scores on the validation set. We truncate the sampling distribution of latents when generating from our models, as previously done in BigGAN~\cite{Brock2018}; we find this to give a slight boost in performance. }
\label{table:results}
\vskip 0.15in
\begin{center}
\begin{small}
\setlength\tabcolsep{5pt}
\begin{tabular}{lcccccc}
\toprule
& & \textsc{train} &  \textsc{train} & \textsc{valid} & \textsc{valid} & \textsc{audio} \\
\textsc{Model} & \textsc{MOS} & \textsc{FDSD} &  \textsc{cFDSD} & \textsc{FDSD} & \textsc{cFDSD} & \textsc{samples} \\
\midrule
\textit{Natural speech} & $4.41 \pm 0.06$ & $0.143$ & & $0.156$ & & \\

\midrule
\textit{Autoregressive models} & & & & \\ 
\textsc{WaveNet} \citep{Kalchbrenner2018} & \phantom{-}$4.51^\dagger\pm0.08$ & & & \\
\textsc{WaveRNN} \citep{Kalchbrenner2018} & \phantom{-}$4.48^\dagger\pm0.07$ & & & \\

\midrule
\textit{Parallel models} & & & & \\
\textsc{MelGAN} \citep{Kumar2019} & \phantom{-}$3.72^\dagger$\hspace{0.95cm} & & & \\
\textsc{Parallel WaveGAN} \citep{Yamamoto2020} & \phantom{-}$4.06^\dagger$\hspace{0.95cm} & & &\\
\textsc{GAN-TTS} \citep{Binkowski2019} & $4.21^\dagger\pm0.05$ & \phantom{-}$0.184$ & \phantom{-}$0.060$ & \\

\midrule
\textit{Our models} & & & & \\ 
\textsc{GAN-TTS} \textit{re-implementation} & $4.16 \pm 0.06$ & $0.163$ & $0.053$ & $0.193$ & $0.077$ & [\href{https://drive.google.com/drive/folders/1coS4cXD-kt8qYRMYVgIsAVYr6KD06Jog?usp=sharing}{Link}] \\
\textsc{GED} \textit{same generator} & $4.03 \pm 0.06$ & $0.151$ & $\bm{0.020}$ & $\bm{0.164}$ & $0.038$ & [\href{https://drive.google.com/drive/folders/1F6xKINmw_-hqY8cJQfA918U4fnY3TKIv?usp=sharing}{Link}] \\
\textsc{GED} \textit{no repulsive term} & $3.00 \pm 0.07$ & $0.145$ & $0.023$ & $0.171$ & $0.048$ & [\href{https://drive.google.com/drive/folders/1V_qidV2LJwqX4ADAfABijo58Xp-m51mV?usp=sharing}{Link}] \\
\textsc{GED} + iSTFT \textit{generator} & $4.10 \pm 0.06$ & $\bm{0.138}$ & $\bm{0.020}$ & $\bm{0.164}$ & $\bm{0.037}$ & [\href{https://drive.google.com/drive/folders/12LRSEJL9HJwqn3wdFU4mKqsFXEw92CWm?usp=sharing}{Link}] \\
\textsc{GED} \textit{+ unconditional} GAN & $\bm{4.25 \pm 0.06}$ & $0.147$ & $0.030$ & $0.169$ & $0.040$ & [\href{https://drive.google.com/drive/folders/1LvFwiZLc-kZRwHsTzIRl2rGNNiOdk7JW?usp=sharing}{Link}] \\
\bottomrule
\end{tabular}

$^\dagger$ Mean Opinion Scores reported by other works are included for reference, but may not be directly comparable due to differences in the data, in the composition of human evaluators, or in the evaluation instructions.
\end{small}
\end{center}
\end{table}

\subsection{Discussion}
\paragraph{Spectral energy distance for TTS}
We studied the effect that switching from adversarial training to training with the spectral energy distance has on the resulting models. To minimize the sources of variation we used a generator architecture similar to that of the GAN-TTS (see Section~\ref{section:model:architecture} and Appendix~\ref{appendix:figure:gan-tts-architecture}). Table~\ref{table:results} shows that in terms of the cFDSD scores models trained with the GED loss improve by $\approx2\times$ on the previously published adversarial model, GAN-TTS, suggesting that they are better at capturing the underlying conditional waveform distributions. We verified that this improvement is not due to overfitting by re-training the models on a smaller dataset ($38.6$~hours) and re-computing these metrics on a larger validation set ($5.8$~hours). We note, however, that the improved cFDSD scores did not transfer to higher subjective speech quality. In fact GED-only samples achieve lower MOS than the adversarial baseline, and empirically we found that FDSD metrics are most informative when comparing different versions of the same architecture and training setup, but not across different models or training procedures.

In two ablation studies (see Appendix~\ref{appendix:section:ablation} and \ref{appendix:section:ablation2}) of the components of our spectral energy loss we confirmed that the GED's repulsive term and the use of multiple scales in the spectral distance function are important for model performance. Moreover, a comparison between the results for \textsc{GED} and \textsc{GED}~\textit{no repulsive term} in Table~\ref{table:results} shows a significant decrease in MOS scores when the repulsive term is not present; and qualitatively, in the absence of the repulsive term, the generated speech sounds metallic. Since using spectrogram-based losses without the repulsive term is standard practice, we feel that comparison against this baseline is most informative in forecasting how useful the proposed techniques will be for the wider community.

\paragraph{Combining GED and adversarial training} The GED loss provides a strong signal for learning the conditional (local) waveform distribution given the linguistic features, but unlike GANs it does not explicitly emphasize accurately capturing the marginal distribution of speech. Empirically, we find that our GED-trained models can sometimes generate audio that, while perfectly audible and closely matching the original speech in timing and intonation, might still sound somewhat robotic. This suggests that these models might still benefit from the addition of an adversarial loss that specifically emphasizes matching the marginal distribution of speech. To test this, we trained the GAN-TTS architecture with its \textit{conditional} discriminators replaced by a single GED loss. The resulting model (\textsc{GED}~+~\textit{unconditional}~\textsc{GAN} in Table~\ref{table:results}) improves on the GED-only model as well as on GAN-TTS, achieving the best-in-class MOS of $4.25 \pm 0.06$.

\paragraph{Choice of network architectures} Encouraged by the stable training of our models with GED, we explored alternative architectures for speech synthesis, like our iSTFT generator (see Section~\ref{section:model:architecture} and Appendix~\ref{appendix:istft}) that generates the coefficients of a Fourier basis and uses them within the inverse STFT transform to produce the final waveform. We find (\textsc{GED}~+~iSTFT~\textit{generator} in Table~\ref{table:results}) that this architecture achieves the best training and validation (c)FDSD scores of the models we tried. In addition, it trains the fastest of our models, reaching optimal cFDSD in as little as 10 thousand parameter updates, with the per-update running time being about half that of the GAN-TTS generator. Unfortunately, this model did not significantly improve upon our results with the simplified GAN-TTS generator in terms of MOS. We tried using the iSTFT architecture in combination with adversarial learning but did not manage to get it to train in a stable way. This supports our belief that the spectral energy distance proposed in this work has the potential to enable the use of a much wider class of network architectures in generative modeling applications, and the design of novel architectures meeting the needs of specific applications (e.g. on-device efficiency).

\paragraph{Train/test performance and overfitting} Our models trained on the generalized energy distance are able to very quickly obtain good cFDSD scores after just 10 to 20 thousand parameter updates. When trained longer without any regularization, validation performance starts to deteriorate after that point. Unregularized, our models are able to produce samples on the training set that are very hard to distinguish from the data. We are actively working on developing new regularization techniques that more effectively translate this capacity into test set performance as measured by MOS.

\section{Conclusion}
We proposed a new generalized energy distance for training generative models of speech without requiring a closed form expression for the data likelihood. Our spectral energy distance is a proper scoring rule with respect to the distribution over spectrograms of the generated waveform audio. The distance can be calculated from minibatches without bias, and does not require adversarial learning, yielding a stable and consistent method for training implicit generative models. Empirical results show that our proposed method is competitive with the state of the art in this model class, and improves on it when combined with adversarial learning.

Our proposed spectral energy distance is closely related to other recent work in audio synthesis~\citep{Wang2019,Yamamoto2020,Engel2020,openai_jukebox}, in that it is based on calculating distances between spectrograms, but we believe it is the first to include a repulsive term between generated samples, and thus the first proper scoring rule of this type. We empirically verified that this is important for obtaining optimal generation quality in our case. Applying our scoring rule to the applications of these other works may offer similar benefits.

With the model and learning method we propose here, we take a step towards closing the performance gap between autoregressive and parallel generative models of speech. With further modeling effort and careful implementation, we hope that our method will be used to enable faster and higher quality generation of audio in live text-to-speech as well as other practical applications.

\section*{Broader impact}
The primary contributions of this paper introduce methodological innovations that improve the automated generation of speech audio from text. Positive aspects of automated text to speech could include improved accessibility for blind and elderly people or others who have poor eyesight.  TTS is a cornerstone of assistive technology and e.g. is already used in the classroom to aid children with developmental disorders with reading comprehension.  Although it is not within the scope of this work, automated TTS could be re-purposed to mimic a specific individual towards benevolent goals (e.g. to comfort someone with the voice of a loved one) or nefarious goals (e.g. to fake someone's voice without their permission).

\section*{Funding disclosure}
This work was funded by Google. None of the authors had financial relationships with other entities relevant to this work.

\section*{Acknowledgments}
We would like to thank Heiga Zen, Norman Casagrande and Sander Dieleman for their insightful comments, help with get acquainted with speech synthesis research and best practices, and for their aid with reproducing GAN-TTS results.

\bibliographystyle{plainnat}
\bibliography{main}

\begin{thebibliography}{39}
\providecommand{\natexlab}[1]{#1}
\providecommand{\url}[1]{\texttt{#1}}
\expandafter\ifx\csname urlstyle\endcsname\relax
  \providecommand{\doi}[1]{doi: #1}\else
  \providecommand{\doi}{doi: \begingroup \urlstyle{rm}\Url}\fi

\bibitem[Amodei et~al.(2016)Amodei, Ananthanarayanan, Anubhai, Bai, Battenberg,
  Case, Casper, Catanzaro, Cheng, Chen, et~al.]{Amodei2016}
Dario Amodei, Sundaram Ananthanarayanan, Rishita Anubhai, Jingliang Bai, Eric
  Battenberg, Carl Case, Jared Casper, Bryan Catanzaro, Qiang Cheng, Guoliang
  Chen, et~al.
\newblock {Deep Speech 2}: {End-to-end} speech recognition in {English} and
  {Mandarin}.
\newblock In \emph{International conference on machine learning}, pages
  173--182, 2016.

\bibitem[Arjovsky et~al.(2017)Arjovsky, Chintala, and Bottou]{Arjovsky2017}
Martin Arjovsky, Soumith Chintala, and L{\'e}on Bottou.
\newblock Wasserstein gan.
\newblock \emph{arXiv preprint arXiv:1701.07875}, 2017.

\bibitem[Bi{\'n}kowski et~al.(2019)Bi{\'n}kowski, Donahue, Dieleman, Clark,
  Elsen, Casagrande, Cobo, and Simonyan]{Binkowski2019}
Miko{\l}aj Bi{\'n}kowski, Jeff Donahue, Sander Dieleman, Aidan Clark, Erich
  Elsen, Norman Casagrande, Luis~C Cobo, and Karen Simonyan.
\newblock High fidelity speech synthesis with adversarial networks.
\newblock \emph{arXiv preprint arXiv:1909.11646}, 2019.

\bibitem[Bouchacourt et~al.(2016)Bouchacourt, Mudigonda, and
  Nowozin]{Bouchacourt2016}
Diane Bouchacourt, Pawan~K Mudigonda, and Sebastian Nowozin.
\newblock Disco nets: Dissimilarity coefficients networks.
\newblock In \emph{Advances in Neural Information Processing Systems}, pages
  352--360, 2016.

\bibitem[Brock et~al.(2018)Brock, Donahue, and Simonyan]{Brock2018}
Andrew Brock, Jeff Donahue, and Karen Simonyan.
\newblock Large scale {GAN} training for high fidelity natural image synthesis.
\newblock \emph{arXiv preprint arXiv:1809.11096}, 2018.

\bibitem[Dhariwal et~al.(2020)Dhariwal, Jun, Payne, Kim, Radford, and
  Sutskever]{openai_jukebox}
Prafulla Dhariwal, Heewoo Jun, Christine Payne, Jong~Wook Kim, Alec Radford,
  and Ilya Sutskever.
\newblock Jukebox: A generative model for music.
\newblock \emph{https://cdn.openai.com/papers/jukebox.pdf}, 2020.

\bibitem[Dinh et~al.(2016)Dinh, Sohl-Dickstein, and Bengio]{Dinh2016}
Laurent Dinh, Jascha Sohl-Dickstein, and Samy Bengio.
\newblock Density estimation using {Real} {NVP}.
\newblock \emph{arXiv preprint arXiv:1605.08803}, 2016.

\bibitem[Dziugaite et~al.(2015)Dziugaite, Roy, and Ghahramani]{Dziugaite2015}
Gintare~Karolina Dziugaite, Daniel~M Roy, and Zoubin Ghahramani.
\newblock Training generative neural networks via maximum mean discrepancy
  optimization.
\newblock \emph{arXiv preprint arXiv:1505.03906}, 2015.

\bibitem[Engel et~al.(2020)Engel, Hantrakul, Gu, and Roberts]{Engel2020}
Jesse Engel, Lamtharn Hantrakul, Chenjie Gu, and Adam Roberts.
\newblock {DDSP}: {Differentiable Digital Signal Processing}.
\newblock \emph{arXiv preprint arXiv:2001.04643}, 2020.

\bibitem[Gneiting and Raftery(2007)]{Gneiting2007}
Tilmann Gneiting and Adrian~E Raftery.
\newblock Strictly proper scoring rules, prediction, and estimation.
\newblock \emph{Journal of the American statistical Association}, 102\penalty0
  (477):\penalty0 359--378, 2007.

\bibitem[Goodfellow et~al.(2014)Goodfellow, Pouget-Abadie, Mirza, Xu,
  Warde-Farley, Ozair, Courville, and Bengio]{Goodfellow2014}
Ian Goodfellow, Jean Pouget-Abadie, Mehdi Mirza, Bing Xu, David Warde-Farley,
  Sherjil Ozair, Aaron Courville, and Yoshua Bengio.
\newblock {Generative} {Adversarial} {Nets}.
\newblock In \emph{Advances in neural information processing systems}, pages
  2672--2680, 2014.

\bibitem[Gretton et~al.(2012)Gretton, Borgwardt, Rasch, Sch{\"o}lkopf, and
  Smola]{Gretton2012}
Arthur Gretton, Karsten~M Borgwardt, Malte~J Rasch, Bernhard Sch{\"o}lkopf, and
  Alexander Smola.
\newblock A kernel two-sample test.
\newblock \emph{Journal of Machine Learning Research}, 13\penalty0
  (Mar):\penalty0 723--773, 2012.

\bibitem[He et~al.(2016)He, Zhang, Ren, and Sun]{He2016}
Kaiming He, Xiangyu Zhang, Shaoqing Ren, and Jian Sun.
\newblock Deep residual learning for image recognition.
\newblock In \emph{Proceedings of the IEEE conference on computer vision and
  pattern recognition}, pages 770--778, 2016.

\bibitem[Heusel et~al.(2017)Heusel, Ramsauer, Unterthiner, Nessler, and
  Hochreiter]{Heusel2017}
Martin Heusel, Hubert Ramsauer, Thomas Unterthiner, Bernhard Nessler, and Sepp
  Hochreiter.
\newblock {GANs} trained by a two time-scale update rule converge to a local
  {Nash} equilibrium.
\newblock In \emph{Advances in neural information processing systems}, pages
  6626--6637, 2017.

\bibitem[Kalchbrenner et~al.(2018)Kalchbrenner, Elsen, Simonyan, Noury,
  Casagrande, Lockhart, Stimberg, {van den Oord}, Dieleman, and
  Kavukcuoglu]{Kalchbrenner2018}
Nal Kalchbrenner, Erich Elsen, Karen Simonyan, Seb Noury, Norman Casagrande,
  Edward Lockhart, Florian Stimberg, Aaron {van den Oord}, Sander Dieleman, and
  Koray Kavukcuoglu.
\newblock Efficient neural audio synthesis.
\newblock \emph{arXiv preprint arXiv:1802.08435}, 2018.

\bibitem[Kim et~al.(2018)Kim, Lee, Song, Kim, and Yoon]{Kim2018}
Sungwon Kim, Sang-gil Lee, Jongyoon Song, Jaehyeon Kim, and Sungroh Yoon.
\newblock {FloWaveNet}: A generative flow for raw audio.
\newblock \emph{arXiv preprint arXiv:1811.02155}, 2018.

\bibitem[Kingma and Ba(2014)]{Kingma2014}
Diederik~P Kingma and Jimmy Ba.
\newblock Adam: {A} method for stochastic optimization.
\newblock \emph{arXiv preprint arXiv:1412.6980}, 2014.

\bibitem[Kingma and Dhariwal(2018)]{Kingma2018}
Durk~P Kingma and Prafulla Dhariwal.
\newblock Glow: {Generative} flow with invertible \verb=1x1= convolutions.
\newblock In \emph{Advances in Neural Information Processing Systems}, pages
  10215--10224, 2018.

\bibitem[Kingma et~al.(2016)Kingma, Salimans, Jozefowicz, Chen, Sutskever, and
  Welling]{Kingma2016}
Durk~P Kingma, Tim Salimans, Rafal Jozefowicz, Xi~Chen, Ilya Sutskever, and Max
  Welling.
\newblock Improved variational inference with {Inverse} {Autoregressive}
  {Flow}.
\newblock In \emph{Advances in neural information processing systems}, pages
  4743--4751, 2016.

\bibitem[Kingsbury et~al.(1998)Kingsbury, Morgan, and Greenberg]{Kingsbury1998}
Brian~ED Kingsbury, Nelson Morgan, and Steven Greenberg.
\newblock Robust speech recognition using the modulation spectrogram.
\newblock \emph{Speech communication}, 25\penalty0 (1-3):\penalty0 117--132,
  1998.

\bibitem[Kuchaiev et~al.(2018)Kuchaiev, Ginsburg, Gitman, Lavrukhin, Case, and
  Micikevicius]{Kuchaiev2018}
Oleksii Kuchaiev, Boris Ginsburg, Igor Gitman, Vitaly Lavrukhin, Carl Case, and
  Paulius Micikevicius.
\newblock {OpenSeq2Seq}: extensible toolkit for distributed and mixed precision
  training of sequence-to-sequence models.
\newblock In \emph{Proceedings of Workshop for NLP Open Source Software
  (NLP-OSS)}, pages 41--46, 2018.

\bibitem[Kumar et~al.(2019)Kumar, Kumar, de~Boissiere, Gestin, Teoh, Sotelo,
  de~Br{\'e}bisson, Bengio, and Courville]{Kumar2019}
Kundan Kumar, Rithesh Kumar, Thibault de~Boissiere, Lucas Gestin, Wei~Zhen
  Teoh, Jose Sotelo, Alexandre de~Br{\'e}bisson, Yoshua Bengio, and Aaron~C
  Courville.
\newblock {MelGAN}: Generative adversarial networks for conditional waveform
  synthesis.
\newblock In \emph{Advances in Neural Information Processing Systems}, pages
  14881--14892, 2019.

\bibitem[Li et~al.(2015)Li, Swersky, and Zemel]{Li2015}
Yujia Li, Kevin Swersky, and Rich Zemel.
\newblock Generative moment matching networks.
\newblock In \emph{International Conference on Machine Learning}, pages
  1718--1727, 2015.

\bibitem[Lim and Ye(2017)]{Lim2017}
Jae~Hyun Lim and Jong~Chul Ye.
\newblock Geometric {GAN}.
\newblock \emph{arXiv preprint arXiv:1705.02894}, 2017.

\bibitem[Miyato et~al.(2018)Miyato, Kataoka, Koyama, and Yoshida]{Miyato2018a}
Takeru Miyato, Toshiki Kataoka, Masanori Koyama, and Yuichi Yoshida.
\newblock Spectral normalization for generative adversarial networks.
\newblock \emph{arXiv preprint arXiv:1802.05957}, 2018.

\bibitem[Pandey and Wang(2019)]{Pandey2019}
Ashutosh Pandey and DeLiang Wang.
\newblock A new framework for {CNN}-based speech enhancement in the time
  domain.
\newblock \emph{IEEE/ACM Transactions on Audio, Speech, and Language
  Processing}, 27\penalty0 (7):\penalty0 1179--1188, 2019.

\bibitem[Ping et~al.(2018)Ping, Peng, and Chen]{Ping2018}
Wei Ping, Kainan Peng, and Jitong Chen.
\newblock {ClariNet}: {Parallel} wave generation in end-to-end
  {Text-to-Speech}.
\newblock \emph{arXiv preprint arXiv:1807.07281}, 2018.

\bibitem[Ping et~al.(2019)Ping, Peng, Zhao, and Song]{Ping2019}
Wei Ping, Kainan Peng, Kexin Zhao, and Zhao Song.
\newblock {WaveFlow}: {A} compact flow-based model for raw audio.
\newblock \emph{arXiv preprint arXiv:1912.01219}, 2019.

\bibitem[Prenger et~al.(2019)Prenger, Valle, and Catanzaro]{Prenger2019}
Ryan Prenger, Rafael Valle, and Bryan Catanzaro.
\newblock {WaveGlow}: A flow-based generative network for speech synthesis.
\newblock In \emph{ICASSP 2019-2019 IEEE International Conference on Acoustics,
  Speech and Signal Processing (ICASSP)}, pages 3617--3621. IEEE, 2019.

\bibitem[Salimans et~al.(2018)Salimans, Metaxas, Zhang, and
  Radford]{Salimans2018}
Tim Salimans, Dimitris Metaxas, Han Zhang, and Alec Radford.
\newblock Improving {GANs} using optimal transport.
\newblock In \emph{6th International Conference on Learning Representations,
  ICLR 2018}, 2018.

\bibitem[Saxe et~al.(2013)Saxe, McClelland, and Ganguli]{Saxe2013}
Andrew~M Saxe, James~L McClelland, and Surya Ganguli.
\newblock Exact solutions to the nonlinear dynamics of learning in deep linear
  neural networks.
\newblock \emph{arXiv preprint arXiv:1312.6120}, 2013.

\bibitem[Sejdinovic et~al.(2013)Sejdinovic, Sriperumbudur, Gretton, and
  Fukumizu]{Sejdinovic2013}
Dino Sejdinovic, Bharath Sriperumbudur, Arthur Gretton, and Kenji Fukumizu.
\newblock Equivalence of distance-based and {RKHS}-based statistics in
  hypothesis testing.
\newblock \emph{The Annals of Statistics}, pages 2263--2291, 2013.

\bibitem[Shen and Vogelstein(2018)]{Shen2018}
Cencheng Shen and Joshua~T Vogelstein.
\newblock The exact equivalence of distance and kernel methods for hypothesis
  testing.
\newblock \emph{arXiv preprint arXiv:1806.05514}, 2018.

\bibitem[Szegedy et~al.(2016)Szegedy, Vanhoucke, Ioffe, Shlens, and
  Wojna]{Szegedy2016}
Christian Szegedy, Vincent Vanhoucke, Sergey Ioffe, Jon Shlens, and Zbigniew
  Wojna.
\newblock Rethinking the {Inception} architecture for computer vision.
\newblock In \emph{Proceedings of the IEEE conference on computer vision and
  pattern recognition}, pages 2818--2826, 2016.

\bibitem[{van den Oord} et~al.(2016){van den Oord}, Dieleman, Zen, Simonyan,
  Vinyals, Graves, Kalchbrenner, Senior, and Kavukcuoglu]{VanDenOord2016}
Aaron {van den Oord}, Sander Dieleman, Heiga Zen, Karen Simonyan, Oriol
  Vinyals, Alex Graves, Nal Kalchbrenner, Andrew Senior, and Koray Kavukcuoglu.
\newblock {WaveNet}: {A} generative model for raw audio.
\newblock \emph{arXiv preprint arXiv:1609.03499}, 2016.

\bibitem[{van den Oord} et~al.(2017){van den Oord}, Li, Babuschkin, Simonyan,
  Vinyals, Kavukcuoglu, Driessche, Lockhart, Cobo, Stimberg,
  et~al.]{VanDenOord2017}
Aaron {van den Oord}, Yazhe Li, Igor Babuschkin, Karen Simonyan, Oriol Vinyals,
  Koray Kavukcuoglu, George van~den Driessche, Edward Lockhart, Luis~C Cobo,
  Florian Stimberg, et~al.
\newblock Parallel {WaveNet}: {Fast} high-fidelity speech synthesis.
\newblock \emph{arXiv preprint arXiv:1711.10433}, 2017.

\bibitem[Wang et~al.(2019)Wang, Takaki, and Yamagishi]{Wang2019}
Xin Wang, Shinji Takaki, and Junichi Yamagishi.
\newblock Neural source-filter-based waveform model for statistical parametric
  speech synthesis.
\newblock In \emph{ICASSP 2019-2019 IEEE International Conference on Acoustics,
  Speech and Signal Processing (ICASSP)}, pages 5916--5920. IEEE, 2019.

\bibitem[Yamamoto et~al.(2020)Yamamoto, Song, and Kim]{Yamamoto2020}
Ryuichi Yamamoto, Eunwoo Song, and Jae-Min Kim.
\newblock Parallel {WaveGAN}: {A} fast waveform generation model based on
  generative adversarial networks with multi-resolution spectrogram.
\newblock In \emph{ICASSP 2020-2020 IEEE International Conference on Acoustics,
  Speech and Signal Processing (ICASSP)}, pages 6199--6203. IEEE, 2020.

\bibitem[Zen et~al.(2009)Zen, Tokuda, and Black]{Zen2009}
Heiga Zen, Keiichi Tokuda, and Alan~W Black.
\newblock Statistical parametric speech synthesis.
\newblock \emph{Speech communication}, 51\penalty0 (11):\penalty0 1039--1064,
  2009.

\end{thebibliography}

\newpage
\begin{appendices}

\section{A proper scoring rule for speech synthesis}
\label{appendix:proper}
A loss function or \emph{scoring rule} $L(q,\bx)$ measures how well a model distribution $q$ fits data $\bx$ drawn from a distribution $p$. Such a scoring rule is called \emph{proper} if its expectation is minimized when $q=p$. If the minimum is also unique, the scoring rule is called \emph{strictly proper}. In the large data limit, a strictly proper scoring rule can uniquely identify the distribution $p$, which means that it can be used as the basis of a statistically consistent learning method.

In Section~\ref{section:spectral-ged} we propose learning implicit generative models of speech by minimizing the \emph{generalized energy score} \citep{Gneiting2007} given by
\begin{equation}
L_{\text{GED}}(q, \bx_{i}) = \E_{\by_{i},\by_{i}' \sim q} 2 d(\bx_{i},\by_{i}) - d(\by_{i},\by_{i}'),
\label{eq:app_ges}
\end{equation}
where $d$ is a distance function over training examples $\bx_{i}$ and generated samples $\by_{i},\by_{i}'$, both of which can be conditioned on a set of features $\bc_{i}$.

In choosing $d()$, we follow the analysis of \citet[Theorem 5.1, Example 5.7]{Gneiting2007}, who study the family of distance functions of the form $d(\bx_{i},\bx_{j})=||\bx_{i}-\bx_{j}||_{\alpha}^{\beta}$ and prove that this choice makes (\ref{eq:app_ges}) a proper scoring rule for learning $p(\bx)$ if $\alpha \in (0,2]$ and $\beta \in (0,\alpha]$. This includes the special cases of L1 and L2 distance, the latter of which they show leads to a \emph{strictly} proper scoring rule.

Given the restrictions set out by this analysis, and building on the domain-specific work of \citet{Engel2020}, we arrived at the following multi-scale spectrogram loss as our choice for $d$:
\begin{equation}
d(\bx_{i},\bx_{j}) = \sum_{k \in [2^6, \ldots, 2^{11}]}\sum_{t} ||\bs^{k}_{t}(\bx_{i}) - \bs^{k}_{t}(\bx_{j})||_{1} + \alpha_{k}||\log \bs^{k}_{t}(\bx_{i}) - \log \bs^{k}_{t}(\bx_{j})||_{2},
\label{eq:app_d}
\end{equation}
where we sum over a geometrically-spaced sequence of STFT window-lengths between 64 and 2048, and where $\bs^{k}_{t}(\bx_{i})$ denotes the $t$-th timeslice of the spectrogram of $\bx_{i}$ with window-length $k$.

Rather than having a single scoring rule (\ref{eq:app_ges}) combined with a multi-scale distance $d()$, we can equivalently rewrite our loss function as a sum over multiple scoring rules, each having a more simple distance function:
\begin{eqnarray}
L_{\text{GED}}(q, \bx_{i}) & = & \sum_{k \in [2^6, \ldots, 2^{11}]}\sum_{t} L^{k,t}_{1}(q, \bx_{i}) + \alpha_{k} L^{k,t}_{2}(q, \bx_{i}) \\
L^{k,t}_{1}(q, \bx_{i}) & = & \E_{\by_{i},\by_{i}' \sim q} 2 ||\bs^{k}_{t}(\bx_{i}) - \bs^{k}_{t}(\by_{i})||_{1} - ||\bs^{k}_{t}(\by_{i}) - \bs^{k}_{t}(\by'_{i})||_{1} \nonumber\\
L^{k,t}_{2}(q, \bx_{i}) & = & \E_{\by_{i},\by_{i}' \sim q} 2 ||\log \bs^{k}_{t}(\bx_{i}) - \log \bs^{k}_{t}(\by_{i})||_{2} - ||\log \bs^{k}_{t}(\by_{i}) - \log \bs^{k}_{t}(\by'_{i})||_{2}.\nonumber
\end{eqnarray}

Here, each of the individual $L^{k,t}_{1}(q, \bx_{i})$ and $L^{k,t}_{2}(q, \bx_{i})$ terms is a proper scoring rule since it uses a L1 or L2 distance with respect to (the $\log$ of) the spectrogram slice $\bs^{k}_{t}(\bx_{i})$. Furthermore, the sum of multiple proper scoring rules is itself a proper scoring rule, and it is strictly proper as long as at least one of the elements in the sum is strictly proper. This means that our combined loss $L_{\text{GED}}(q, \bx_{i})$ is indeed a strictly proper scoring rule with respect to $p(\bs^{k}_{t})$. It follows that it is also a proper scoring rule for $p(\bx|\bc)$, but not necessarily a strictly proper one, since $\bx$ may have long-range dependencies that cannot be identified from single spectrogram slices $\bs^{k}_{t}$. We also experimented with adding such longer range terms to our training loss but found no additional empirical benefit.

We experimented with various weights $\alpha_{k}$ for the L2 term in (\ref{eq:app_d}), and found $\alpha_{k} = \sqrt{k/2}$ to work well. This choice approximately equalizes the influence of all the different L1 and L2 terms on the gradient with respect to our generator parameters $\theta$. Dropping the L2 terms by setting $\alpha_{k}=0$ only gave us slightly worse results, and could be used as a simpler alternative.

For the calculation of the spectrograms $\bs^{k}(\bx_{i})$ we obtained slightly better sounding results when mapping raw STFT outputs to the \emph{mel}-frequency-scale, but with slightly worse results in terms of cFDSD. All reported results are with mel-scale spectrograms.

\section{Ablation study on the spectral energy distance}
\label{appendix:section:ablation}
We carried out an ablation study, in which we systematically varied aspects of the spectral energy distance proposed in Section~\ref{section:spectral-ged} while using the architecture described in Section~\ref{section:model:architecture}. The results of these ablations are presented in Table~\ref{appendix:table:results-ablations}. We note that at a high level we observe that any deviation from the proposed spectral energy distance leads to higher (worse) values of the validation (c)FDSD metrics, and discuss specific ablation experiments below.

Compared to the Baseline~GED model, the same model \textit{without} the repulsive term (``Generalised energy distance: disabled'' in Table~\ref{appendix:table:results-ablations}) not only gets worse FDSD scores, but also significantly reduces quality of the synthesized speech (see Section~\ref{section:experiments}), suggesting that this part of the loss is crucial for the models ability to accurately capture the underlying conditional waveform distributions.

We compute spectrograms using an overcomplete basis of sinusoids. An exploration of the effect of this oversampling (``DCT / DST overcompleteness'' in Table~\ref{appendix:table:results-ablations}) shows that the FDSD metric values stops improving beyond the use of an $8\times$ overcomplete basis. Another benefit of an overcomplete basis that is not captured by Table~\ref{appendix:table:results-ablations} is faster convergence of models with a more overcomplete basis; but this improvement too tapered off once the basis was at least $8\times$ overcomplete.

Finally, we explored the importance of using a multiple spectrogram scales in the GED loss (``Window sizes'' in Table~\ref{appendix:table:results-ablations}) by training models that each used only a single window size $k$ for its spectrograms. Our results show that individually all of the constituent window sizes yield worse results than when they are combined in a single loss, suggesting that use of multiple spectrogram scales is an important aspect of the proposed spectral energy distance.

\begin{table}[ht]
\caption{Validation FDSD metric values for experiments comparing the proposed model and its variants. The ablation experiments only ran for $200\times10^{3}$ training steps and not until convergence.}
\label{appendix:table:results-ablations}
\vskip 0.15in
\begin{center}
\begin{small}
\begin{tabular}{lrcc}
\toprule
& & \textsc{valid} & \textsc{valid} \\
\textsc{Study} & \textsc{Variant} & \textsc{FDSD} & \textsc{cFDSD} \\
\midrule
\vspace{0.6ex} Baseline GED & & $0.163$ & $0.040$ \\

\vspace{0.6ex} Generalized energy distance & disabled  & $0.170$ & $0.047$ \\

DCT / DST overcompleteness & 1x & $0.165$ & $0.042$ \\
                      & 2x & $0.165$ & $0.041$ \\
                      & 4x & $0.168$ & $0.041$ \\
\vspace{0.6ex}        & 16x & $0.163$ & $0.041$ \\

Window sizes & $64$ & $0.195$ & $0.087$ \\
            & $128$ & $0.168$ & $0.046$ \\
            & $256$ & $0.166$ & $0.043$  \\
            & $512$ & $0.174$ & $0.048$ \\
            & $1024$ & $0.182$ & $0.064$ \\
            & $2048$ & $0.202$ & $0.093$ \\

\bottomrule
\end{tabular}

\end{small}
\end{center}
\end{table}

\section{Ablation study combining GED and GANs}
\label{appendix:section:ablation2}
On the suggestion of the reviewers we performed an additional ablation study to more carefully examine the interaction of an adversarial loss with our proposed GED loss. Table~\ref{tb:gedgan} shows cFDSD and MOS results for all combinations of 1) using a repulsive term or not, 2) using a multi-scale or single-scale spectrogram loss, and 3) using an unconditional GAN loss or not. These experiments ran for the full $10^{6}$ training steps, and include MOS scores as well as cFDSD scores, making them complimentary to the ablation study shown in Table~\ref{appendix:table:results-ablations}.

The results in Table~\ref{tb:gedgan} confirm that including the repulsive term in the spectral energy distance always improves over the naive spectrogram loss in terms of MOS. Furthermore, we find that adding the adversarial loss is generally helpful, and that the multi-scale loss outperforms the single-scale loss. 

\begin{table}[h]

\caption{Results for all combinations of (1) repulsive term (\textbf{r}) yes/no, (2) multi-scale (\textbf{m}) or single window size (256/512) or no spectrogram loss, (3) unconditional GAN loss (\textbf{G}) yes/no.  Note that these ablations sampled the cFDSD validation set uniformly, where we used length-weighted sampling for the main paper and Table~\ref{tb:ged_full_gan} below.}
\label{tb:gedgan}
\vspace{0.2cm}
\resizebox{\textwidth}{!}{
\begin{tabular}{lccccccc}
model$\rightarrow$ & \text{\textbf{r}+\textbf{m}+\textbf{G}} & \text{\textbf{r}+\textbf{m}} &  \text{\textbf{r}+256+\textbf{G}} & \text{\textbf{r}+512+\textbf{G}}  & \text{\textbf{r}+256} & \text{\textbf{r}+512} \\
cFDSD$\rightarrow$ & $0.033$ & $0.033$ & $0.344$ & $0.063$ & $0.035$ & $0.034$ &\\
MOS$\rightarrow$ & $4.25 \pm 0.06$  & $4.06 \pm 0.06$ & $3.67 \pm 0.07$  & $3.96 \pm 0.059$ & $3.44 \pm 0.07$ & $2.89 \pm 0.09$ \\\hline
model$\rightarrow$ & \text{\textbf{m}+\textbf{G}} & \text{\textbf{m}} & \text{256+\textbf{G}} & \text{512+\textbf{G}}  & \text{256} & \text{512}\\ 
cFDSD$\rightarrow$ & $0.039$ & $0.039$ & $0.200$ & $0.047$ & $0.040$ & $0.038$\\ 
MOS$\rightarrow$ & $4.12 \pm 0.06$ & $3.00 \pm 0.07$ &  $2.86 \pm 0.07$ & $3.82 \pm 0.06$ & $2.33 \pm 0.06$ & $2.48 \pm 0.06$ \\\hline 
\end{tabular}}
\end{table}

Finally, we also ran an experiment combining our GED loss with GAN-TTS, with the conditional discriminators of GAN-TTS included. This experiment can be compared against the results in the main paper that only include \emph{unconditional} discriminators when combining GED and GAN. As Table~\ref{tb:ged_full_gan} shows, the combination of full GAN-TTS and GED performs about equally well as our proposed combination of GED and unconditional GAN. Both outperform the baseline of GAN-TTS without GED loss.

\begin{table}[h]
\caption{Results for combining our proposed GED loss with full GAN-TTS, including the conditional discriminators, and comparing against GED + unconditional GAN, and GAN-TTS.}
\label{tb:ged_full_gan}
\vspace{0.2cm}
\centering
\begin{tabular}{lccc}
model$\rightarrow$ & \text{GED + full GAN-TTS} & \text{GED + uncond. GAN} & \text{GAN-TTS only}\\
cFDSD$\rightarrow$ & $0.041$ & $0.040$ & $0.077$\\
MOS$\rightarrow$ & $4.24 \pm 0.05$ & $4.25 \pm 0.06$ & $4.16 \pm 0.06$ \\\hline
\end{tabular}
\end{table}

\section{Training and architecture details}

\subsection{Spectral distance}
In practice, when computing the STFT spectrograms necessary for the spectral GED loss \eqref{eq:combined_dist}, we found that the training was more stable when spectrograms $s^k_i$ and $s^k_j$ were computed with Hann windowing, $50\%$ overlap and using an \textit{overcomplete} Fourier basis. This is equivalent to transforming the windows of length $k$ using the Discrete Cosine and Discrete Sine (DCT and DST) with basis functions $\cos(\frac{2\pi}{k} \cdot \frac{i}{m})$ and $\sin(\frac{2\pi}{k} \cdot \frac{i}{m})$  to obtain the real and imaginary parts of the STFT, where $m$ is an integer oversampling multiplier and $i=0,\ldots,\frac{mk}{2} + 1$. For $m=1$ this is equivalent to the standard Fourier transform, and we used $m=8$ in our experiments. Importantly, we observed that using an $\times8$ overcomplete basis did not significantly slow down training on modern deep learning hardware.

\subsection{Training details}
Unless otherwise specified, all models were trained with the same hyper-parameters (see Table~\ref{appendix:table:spectre-hyperparamters}) on Cloud TPUs v3 with 128-way data parallelism and cross-replica Batch Normalization. Furthermore, unless otherwise specified, no additional regularization was used, i.e. the spectral energy distance was minimized directly. A single experiment took between 2 and 4 days to complete $10^6$ training steps.

GED~+~\textit{unconditional}~GAN used GAN-TTS hyper-parameters from Table~\ref{appendix:table:gan-tts-hyperparamters}, but with the generator learning rate set to $1\times10^{-4}$. The weight of the GED loss was set to $3$.

GED~+~iSTFT~\textit{generator} used the Adamax \citep{Kingma2014} optimizer with $\beta_1=0.9$, $\beta_2=0.9999$, learning rate $10^{-3}$ with a linear warmup over $12000$ steps, EMA decay rate of $0.99998$ and early stopping to avoid overfitting.

\begin{table}[htb!]
\caption{Default hyper-parameters.}
\label{appendix:table:spectre-hyperparamters}
\vskip 0.15in
\begin{center}
\begin{small}
\begin{tabular}{lr}
\toprule
\textsc{Hyper-parameter} & \textsc{Value} \\

\midrule

Optimizer & Adam \citep{Kingma2014} \\
Adam $\beta_1$ & $0.9$ \\
Adam $\beta_2$ & $0.999$ \\
Adam $\epsilon$ & $10^{-8}$ \\
Learning rate & $3 \times 10^{-4}$ \\
Learning rate schedule & Linear warmup over $6000$ steps \\
Initialization: shortcut convolutions & Zeros \\
Initialization: conditional Batch Normalization & Zeros \\
Initialization: rest & Orthogonal \citep{Saxe2013} \\
EMA decay rate & $0.9999$ \\
Batch Normalization $\epsilon$ & $10^{-4}$ \\
Batch size & $1024$ \\
Training steps & $10^{6}$ \\

\bottomrule
\end{tabular}
\end{small}
\end{center}
\end{table}

\begin{figure}[htb]

\begin{centering}
\begin{subfigure}[t]{0.29\textwidth}
\centering
\includegraphics[scale=0.75]{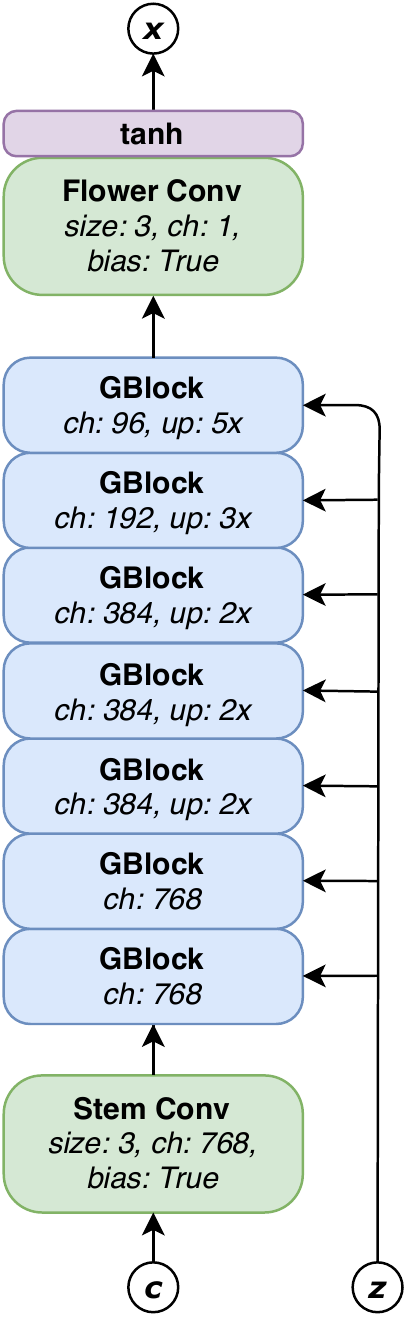}
\caption{Generator architecture.}
\end{subfigure}%
~\hspace{1.2cm}~
\begin{subfigure}[t]{0.69\textwidth}
\centering
\includegraphics[scale=0.75]{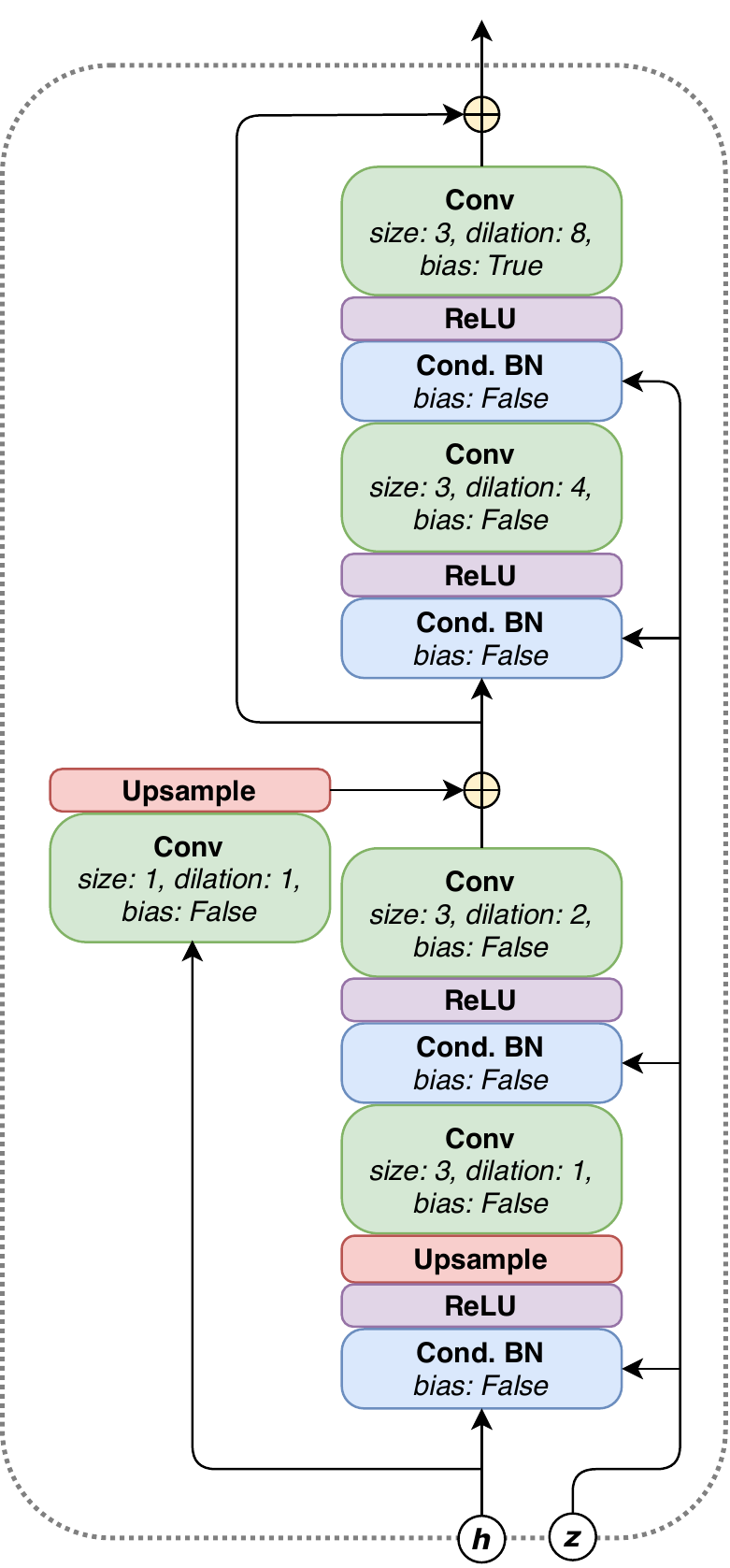}
\caption{Generator residual block (GBlock).}
\end{subfigure}
\end{centering}

\caption{The proposed generative model (\textbf{a}) resembles the GAN-TTS generator and consists of $7$ GBlocks (\textbf{b}) that use convolutional layers of increasing dilation rates, nearest-neighbour upsampling and conditional Batch Normalization. The number of channels is changed only in the block's first and shortcut convolutions; and the latter is only present if the block reduces the number of channels. The residual blocks follow the same upsampling pattern as GAN-TTS.}
\label{fig:spectre-architecture}
\end{figure}

\subsection{Simplified GAN-TTS generator}
\label{appendix:arch}
To convincingly demonstrate the usefulness of the generalized energy distance for learning implicit generative models of speech, we sought to compare it to GAN-TTS, a state-of-the-art adversarial TTS model. To this end in our core experiments we used an architecture that is nearly identical to that of the GAN-TTS generator, but is further simplified as described in Section~\ref{section:model:architecture} and as depicted in Figure~\ref{fig:spectre-architecture}.

\begin{figure}[t]

\begin{centering}
\begin{subfigure}[t]{0.22\textwidth}
\centering
\includegraphics[scale=0.75]{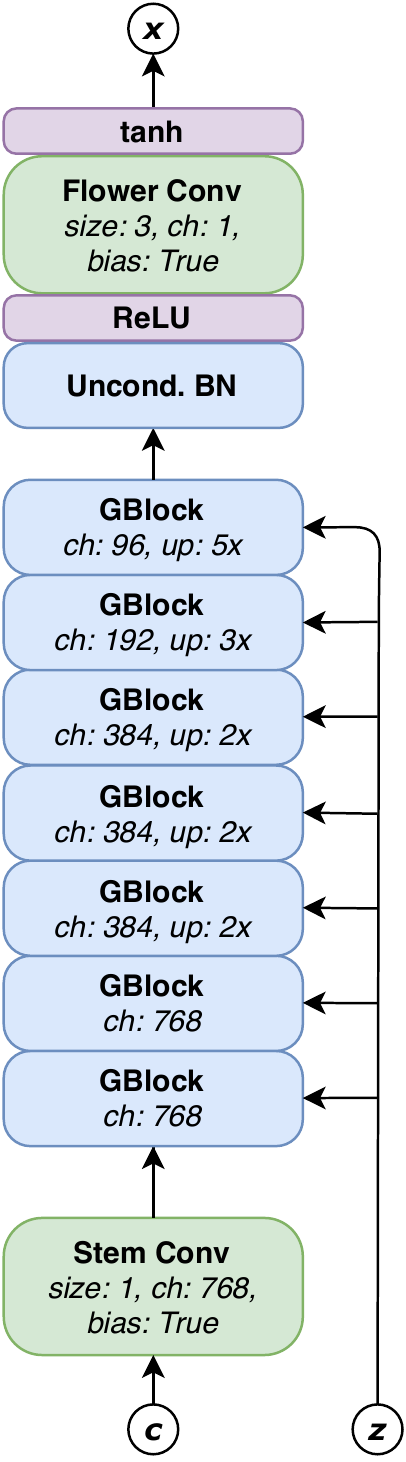}
\caption{Generator.}
\label{appendix:figure:gan-tts-architecture:generator}
\end{subfigure}%
~ 
\begin{subfigure}[t]{0.28\textwidth}
\centering
\includegraphics[scale=0.75]{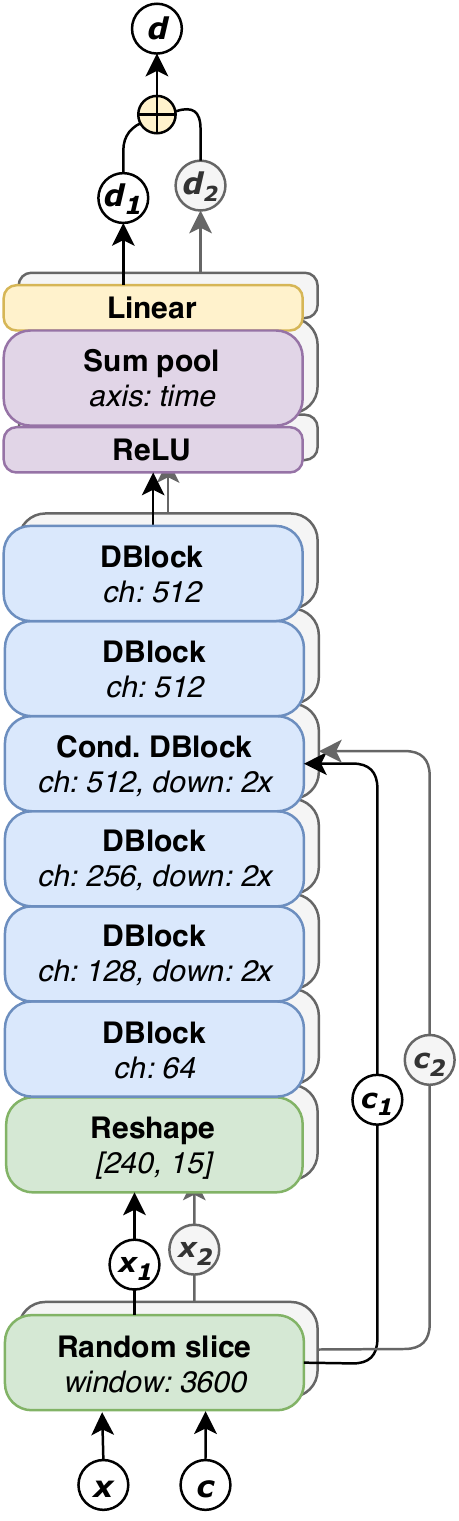}
\caption{Conditional discriminator.}
\label{appendix:figure:gan-tts-architecture:discriminator}
\end{subfigure}
~ 
\begin{subfigure}[t]{0.45\textwidth}
\centering
\includegraphics[scale=0.75]{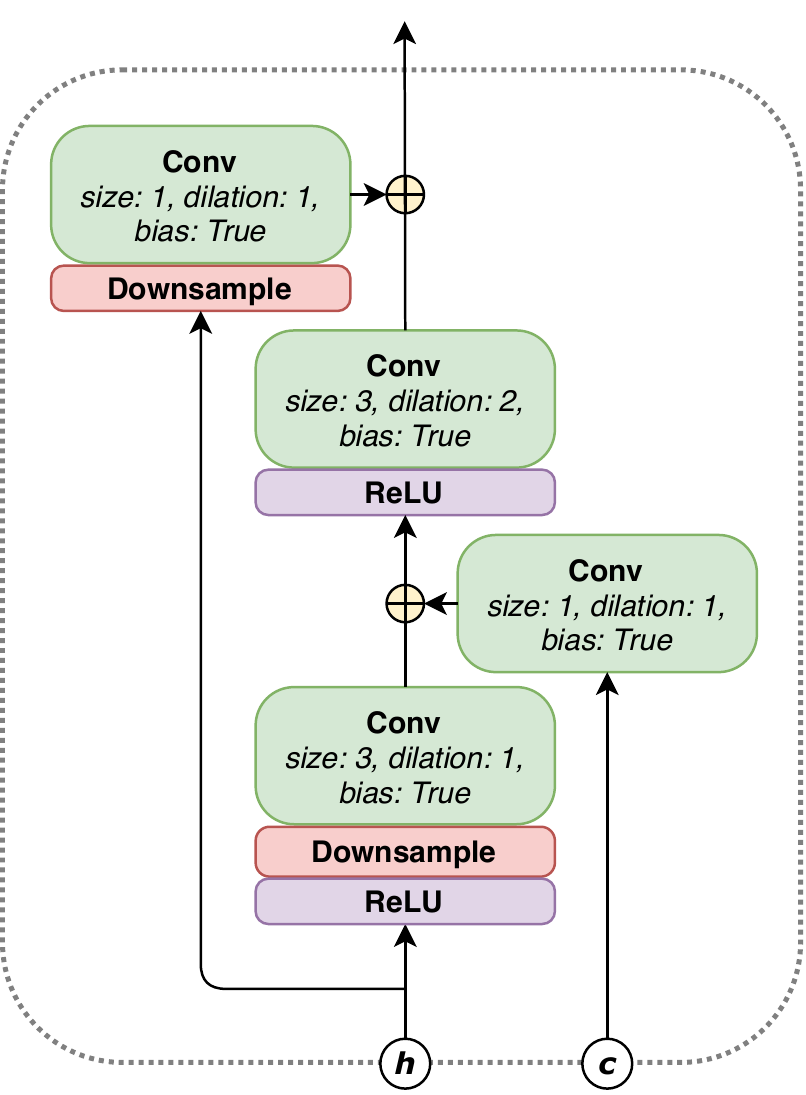}
\caption{Conditional discriminator residual block (DBlock).}
\label{appendix:figure:gan-tts-architecture:dblock}
\end{subfigure}
\end{centering}

\caption{Architectures used in our implementation of GAN-TTS. Generator (\textbf{a}) makes use of a smaller kernel size $1$ convolution in the stem embedding linguistic features $\bc$ and GBlocks identical to those in Figure~\ref{fig:spectre-architecture}. Discriminator (\textbf{b}) replaces mean-pooling with an additional non-linearity, sum-pooling and a final projection layer to obtain the scalars $d_{1,2}$ for each of the two random slices it samples. The random slice block takes aligned random crops (same for every example in the minibatch) $\vect{x}_{1,2}$ and $\bc_{1,2}$ of the waveform $\vect{x}$ and conditioning features $\bc$, and the outputs for each of the two slices are averaged to obtain the final output $d = \frac{1}{2}(d_1+d_2)$. Example architecture is shown for a conditional discriminator with window size $3600$, but the same changes are applied to other window sizes and unconditional discriminators. The modified (conditional) DBlock (\textbf{c}) re-orders the first non-linearity and downsampling blocks.}
\label{appendix:figure:gan-tts-architecture}
\end{figure}

\pagebreak
\subsection{Inverse STFT generator}
\label{appendix:istft}
\begin{wrapfigure}{r}{0.25\textwidth}
  \centering
  \vspace{-40pt}
    \includegraphics[width=0.25\textwidth]{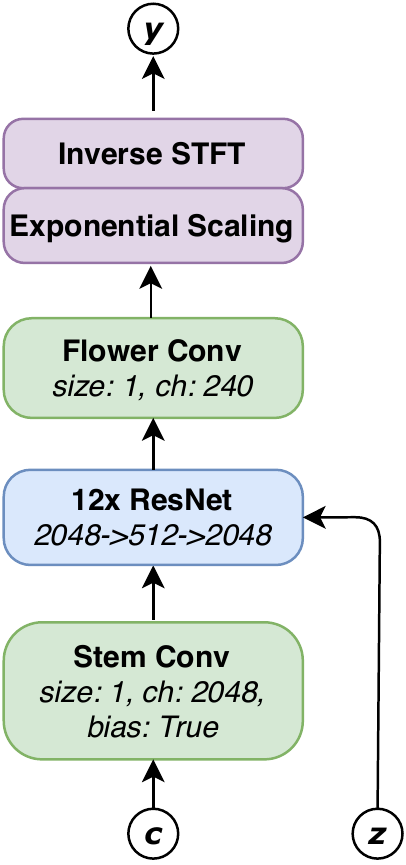}
    \caption{iSTFT model.}
    \label{fig:istft_model}
    \vspace{-45pt}
\end{wrapfigure}
Our inverse STFT generator takes in linguistic features $\bc$ at a frequency of 1 feature vector per 120 timesteps (a \emph{chunk}). A 1D convolution with kernel size 1 is used to project the features to a 2048 dimensional vector per chunk, which is then fed into a stack of 12 \emph{bottleneck} ResNet blocks \citep{He2016}.

Each of the ResNet blocks consists of a kernel size 1 convolution to 512 channels, 2 convolutions of kernel size 5 at 512 channels, followed by projection to 2048 channels again. In-between the convolutions we use conditional batch normalization as also used in GAN-TTS and as described in Section~\ref{appendix:gan-tts}.

Finally we project down to 240 dimensions per chunk. Of these dimensions, one is used to exponentially scale the remaining 239 features. These remaining features are then interpreted as the non-redundant elements of an STFT with window size 240 and frame step 120, and are projected to the waveform space using a linear inverse STFT transformation. The model stack is visualized in Figure~\ref{fig:istft_model}.

\section{GAN-TTS baseline}
\label{appendix:gan-tts}

We re-implemented the GAN-TTS model from \citet{Binkowski2019} for use as a baseline in our experiments. While attempting to reproduce the original implementation as closely as possible by following the description provided in \citep{Binkowski2019}, we observed that our implementation of the model (\textbf{i}) would not match the reported FDSD scores (reaching an cFDSD of $\approx2.5$ instead of the reported $0.06$); and (\textbf{ii}) would diverge during training. To alleviate these discrepancies, we found it necessary to deviate from the architecture and training procedure described in \citet{Binkowski2019} in several ways detailed below. Our modified implementation reaches cFDSD of $0.056$ and trains stably.

\begin{figure}[ht]
\begin{centering}
\includegraphics[width=\textwidth, trim=3cm 1cm 3.5cm 1cm]{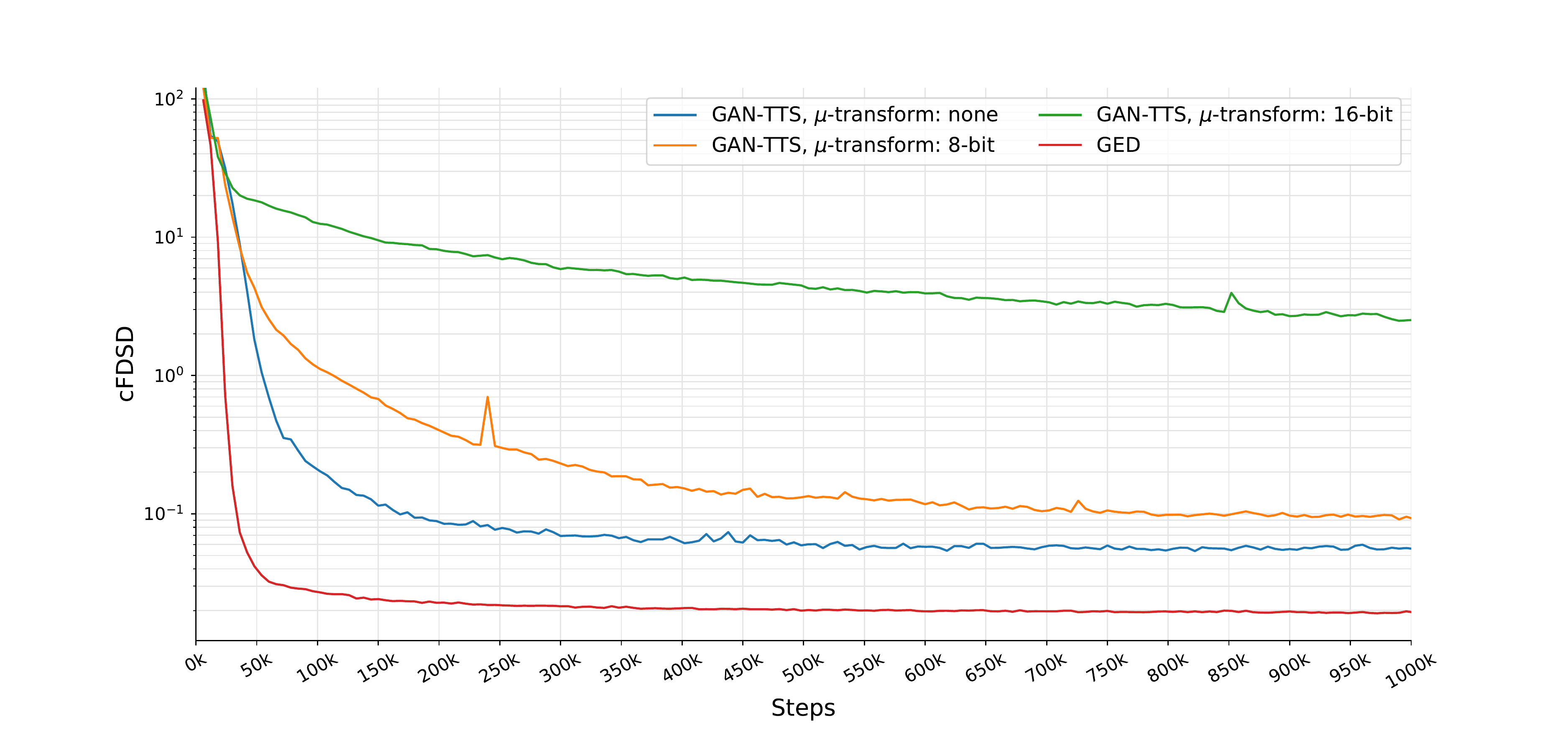}
\end{centering}

\caption{GAN-TTS baseline (\textbf{our implementation}) with and without $\mu$-transform; and GED convergence speed and training stability.}
\label{fig:gan-tts-vs-spectre}
\end{figure}

\paragraph{No $\mu$-transform.}
We found that the use of a $\mu$-transform with 16-bit encoding ($\mu=2^{16}-1$) was the single largest factor responsible for low-quality samples in our initial implementation of GAN-TTS. With the $\mu$-transform enabled (i.e. generating and discriminating transformed audio), our GAN-TTS baseline converged very slowly and would only reach cFDSD of $\approx2.5$ (see Figure~\ref{fig:gan-tts-vs-spectre}). Disabling the $\mu$-transform was necessary for reaching competitive sample quality ($0.056$ cFDSD and $4.16$ MOS). We also observed that the use of $\mu$-transform made training more unstable.

\paragraph{Generator architecture.}
We re-used most of the original generator architecture described in \citet{Binkowski2019}, but empirically found that (\textbf{i}) adding a batch normalization followed by a non-linearity before the flower convolution; and (\textbf{ii}) switching from a kernel size $3$ to a kernel size $1$ convolution; both led to more stable training with default settings. Addition of the former is inspired by the BigGAN architecture \citep{Brock2018} that GAN-TTS is based on; and the latter relies on an interpretation of the first convolution as an embedding layer for the sparse conditioning linguistic features. These differences are reflected in the generator architecture in Figure~\ref{appendix:figure:gan-tts-architecture:generator}.

\paragraph{Discriminator architecture.}
Empirically we found that it was necessary to introduce more changes to the discriminator architecture. Specifically, the following alterations were made (see also Figure~\ref{appendix:figure:gan-tts-architecture:discriminator} and Figure~\ref{appendix:figure:gan-tts-architecture:dblock}):
\begin{itemize}
    \item The mean-pooling along time and channel axes of the output of the final DBlock was replaced by a non-linearity, followed by sum-pooling along the time axis and a dense linear projection to obtain a scalar output. Like the addition of batch normalization and non-linearity before the generator output, this change is inspired by the BigGAN architecture.
    
    \item Instead of a single random slice of the input waveforms, each discriminator sampled \textit{two} random slices $(\bx_1, \bc_1)$ and $(\bx_2, \bc_2)$, and produced independent outputs $d_1$ and $d_2$ for each of them. These outputs were later averaged to produce discriminators final output $d$.
    
    \item Inspired by the open source implementation of BigGAN\footnote{See \url{https://github.com/ajbrock/BigGAN-PyTorch}}, the structure of the first DBlock of each discriminator was altered to not include the first non-linearity. The architecture was surprisingly sensitive to this detail.
    
    \item Finally, the structure of the DBlocks was modified by (\textbf{i}) switching the order of the downsampling and non-linearity operations; and (\textbf{ii}) by reducing the dilation of the second convolution to $1$ when the time dimension of the block is less or equal to $16$.
\end{itemize}

\paragraph{Hyper-parameters.}
\begin{table}[htb]
\caption{Hyper-parameters used in our implementation of the GAN-TTS baseline.}
\label{appendix:table:gan-tts-hyperparamters}
\vskip 0.15in
\begin{center}
\begin{small}
\begin{tabular}{lr}
\toprule
\textsc{Hyper-parameter} & \textsc{Value} \\

\midrule

Optimizer & Adam \citep{Kingma2014} \\
Adam $\beta_1$ & $0$ \\
Adam $\beta_2$ & $0.999$ \\
Adam $\epsilon$ & $10^{-6}$ \\
Generator learning rate & $5 \times 10^{-5}$ \\
Discriminator learning rate & $10^{-4}$ \\
Learning rate schedule & Linear warmup over $6000$ steps \\
Loss & Hinge \citep{Lim2017} \\
Initialization & Orthogonal \citep{Saxe2013} \\
Generator EMA decay rate & $0.9999$ \\
Batch Normalization $\epsilon$ & $10^{-4}$ \\
Batch Normalization momentum & $0.99$ \\
Spectral Normalization $\epsilon$ & $10^{-4}$ \\
Batch size & $1024$ \\
Training steps & $10^{6}$ \\

\bottomrule
\end{tabular}
\end{small}
\end{center}
\end{table}
We recap all hyper-parameters used in our re-implementation of GAN-TTS in Table~\ref{appendix:table:gan-tts-hyperparamters}. As in the original publication, the GAN-TTS baseline was trained on a Cloud TPUs v3 with 128-way data parallelism and cross-replica Batch Normalization; training a single model took approximately $48$~hours.

Training curves for our implementation of GAN-TTS, and how they compare to a similar (simplified) generator trained with the GED loss is shown in Figure~\ref{fig:gan-tts-vs-spectre}.

\subsection{Fr\'echet Deep Speech Distances}
\label{appendix:FDSD}
At the time of writing no open source implementation of the Fr\'echet Deep Speech Distance (FDSD) metrics \citep{Binkowski2019} was available. We thus resorted to re-implementating these metrics based on the information provided in the original publication. While striving to reproduce the original implementation as closely as possible, we deviated from it in at least two aspects, as discussed below.

Following the notation of \citet{Binkowski2019}, let $\vect{a}\in\mathbb{R}^{48000}$ be a vector representing two seconds of (synthesized) waveform at 24 kHz; $\text{DS}(\vect{a}) \in \mathbb{R}^{1600}$ be the sought representation that will be used for computing the (conditional) FDSD; and $f_{k\omega}: \mathbb{R}^{k\omega} \mapsto \mathbb{R}^{\lceil\frac{k}{2}\rceil \times 1600}$ be a function that takes (a part of) the waveform $\vect{a}$ and passes it through the pre-trained Deep~Speech~2 (DS2) network \citep{Amodei2016, Kuchaiev2018} to obtain the necessary activations. The representation $\text{DS}(\vect{a})$ used for computing the Fr\'echet distance is then obtained by averaging the outputs of $f$ across time.

\begin{enumerate}
    \item Equation~(4) in Appendix~B.1 of \citet{Binkowski2019} implies that the necessary activations were obtained \emph{independently} for windows of the waveform $\vect{a}$ with window size $\omega=480$ and step $\frac{\omega}{2}=240$ (20ms and 10ms at 24kHz respectively), resulting $199$ activation vectors of size $1600$ each, which were then averaged to obtain the representation $\text{DS}(\vect{a})$. Doing so would not make any use of the DS2 bi-directional GRU layers, as their inputs would have time dimensionality of $1$ - a single frame of the STFT with frame length $\omega$ and step $\frac{\omega}{2}$. So instead we used the entire audio fragment $\vect{a}$ ($200$ STFT frames) at once to obtain activations $f_{48000}(\vect{a})\in\mathbb{R}^{100\times1600}$ that were averaged along the time axis to obtain $\text{DS}(\vect{a})$.

    \item \citet{Binkowski2019} proposed using activations from the node labeled \verb=ForwardPass/ds2_encoder/Reshape_2= in the graph of a pre-trained DS2 network to obtain activations $f_{k\omega}$. This graph node belongs to the training pass of the model, and uses $6$ layers with $0.5$ dropout probability (one after each of the $5$ GRU layers, and then again after the last fully-connected layer of the encoder network), resulting very sparse activations. To make better use of the learned representations, we instead used the graph node labeled \verb=ForwardPass_1/ds2_encoder/Reshape_2=, which implements the test time behaviour of the same network and produces dense activations.
\end{enumerate}

The rest of the implementation followed \citet{Binkowski2019}. Namely, FDSD were estimated using $10000$ samples from the \textit{training} data, matching the conditioning signals between the two sets in the case of conditional FDSD (cFDSD).

We tested our implementation by computing the FDSD for natural speech - the only quantity from \citet{Binkowski2019} that can be reproduced without access to a trained generator, and found that despite the implementation differences it agrees surprisingly well with the previously reported number (ours: $0.143$ vs. \citet{Binkowski2019}: $0.161$). We also considered implementations of the FDSD that did not deviate from the original description (i.e. using dropout and/or obtaining activations for each window independently), but found that they had worse agreement with the previously reported natural speech FDSD.

Without access to the original implementation it is impossible to tell whether there are other differences between the two FDSD implementations, or whether the described differences are actually there - the two implementations agree unexpectedly well on natural speech FDSD despite significant discrepancies in how they extract representations from the pre-trained model.

We hope that the difficulties we faced reproducing these results will prompt the research community to open-source evaluation metrics early on, even in cases when the models themselves cannot be made publicly available. We provide our implementation of FDSD in our github repository at \url{https://github.com/google-research/google-research/tree/master/ged_tts}.

\section{Mean Opinion Scores}
Each evaluator, a native North American English speaker paid to perform the task was asked to rate the subjective naturalness of a sentence on a 1-5 (Bad-Excellent) Likert scale. Mean Opinion Scores (MOS) were obtained by summarizing as mean and standard deviation the $1000$ audio sample ratings produced by at least $80$ different human evaluators per test. The resulting scores are comparable between between the models trained in this work, but may not be directly comparable with previous work due to differences in composition of human evaluators and the evaluation instructions given to them.

\section{Linguistic features}
As in \citep{Binkowski2019, Kalchbrenner2018, VanDenOord2016, VanDenOord2017}, synthesized speech was conditioned on local linguistic features and pitch information \textit{predicted} from text using separate models; and ground truth linguistic features and pitch were used during training.

\end{appendices}

\end{document}